\documentclass[preprint]{acmart}

\AtBeginDocument{%
  \providecommand\BibTeX{{%
    \normalfont B\kern-0.5em{\scshape i\kern-0.25em b}\kern-0.8em\TeX}}}

\usepackage[T1]{fontenc}
\usepackage[utf8]{inputenc}
\usepackage{lmodern}
\usepackage{graphicx}
\usepackage{hyperref}
\usepackage{csquotes}
\usepackage[english]{babel}
\setcopyright{none}
\renewcommand\footnotetextcopyrightpermission[1]{} % removes footnote with conference information in first column
\begin{document}

\title{Mitigating Bias in Algorithmic Systems - A Fish-Eye View}
%\title{Fairness and Transparency in Algorithmic Systems: \\ An Integrated Framework}

\author{Kalia Orphanou}
\email{kalia.orphanou@ouc.ac.cy}
\orcid{1234-5678-9012}
\affiliation{%
  \institution{Open University of Cyprus}
  %\streetaddress{P.O. Box 1212}
\country{CYPRUS}
}

\author{Jahna Otterbacher}
\email{jahna.otterbacher@ouc.ac.cy}
\author{Styliani Kleanthous}
%\email{styliani.kleanthous@ouc.ac.cy}
\affiliation{%
  \institution{Open University of Cyprus \& \\ CYENS Centre of Excellence}
 % \streetaddress{P.O. Box 1212}
\country{CYPRUS}
}

\author{Khuyagbaatar Batsuren}
\authornote{Work conducted while at The University of Trento.}
\affiliation{%
  \institution{National University of Mongolia}
  \city{Ulan Baatar}
  \country{MONGOLIA}}
%\email{k.batsuren@unitn.it}

\author{Fausto Giunchiglia}
\affiliation{%
  \institution{The University of Trento}
  \streetaddress{1 Th{\o}rv{\"a}ld Circle}
  \city{Trento}
  \country{ITALY}}
%\email{larst@affiliation.org}

\author{Veronika Bogina}
\author{Avital Shulner Tal}
%\email{webmaster@marysville-ohio.com}
\author{Alan Hartman}
%\email{webmaster@marysville-ohio.com}
\author{Tsvi Kuflik}
\affiliation{%
  \institution{The University of Haifa}
 % \streetaddress{P.O. Box 1212}
\country{ISRAEL}
}

%\affiliation{%
% \institution{Rajiv Gandhi University}

%\affiliation{%
%  \institution{Tsinghua University}
%  \streetaddress{30 Shuangqing Rd}

%\affiliation{%
 % \institution{Palmer Research Laboratories}
 % \streetaddress{8600 Datapoint Drive}
 % \city{San Antonio}
 % \state{Texas}
 % \postcode{78229}}

%\affiliation{\institution{The Th{\o}rv{\"a}ld Group}}
%\email{jsmith@affiliation.org}

%\affiliation{\institution{The Kumquat Consortium}}
%\email{jpkumquat@consortium.net}

%\usepackage{longtable}

\renewcommand{\shortauthors}{Orphanou and Otterbacher, et al.}

\begin{abstract}
Mitigating bias in algorithmic systems is a critical issue drawing attention across communities within the information and computer sciences. Given the complexity of the problem and the involvement of multiple stakeholders -- including developers, end users and third-parties -- there is a need to understand the landscape of the sources of bias, and the solutions being proposed to address them, from a broad, cross-domain perspective. This survey provides a ``fish-eye view,'' examining approaches across four areas of research. The literature describes three steps toward a comprehensive treatment -- bias detection, fairness management and explainability management -- and underscores the need to work from within the system as well as from the perspective of stakeholders in the broader context.  
\end{abstract}




\maketitle

\section{Introduction}
Long before the widespread use of algorithmic systems driven by big data, Friedman and Nissenbaum \cite{friedman1996bias}, writing in the ACM TOIS in 1996, argued that “freedom from bias” should be considered equally alongside the criteria of reliability, accuracy and efficiency, when judging the quality of a computer system. Defining biased systems as those that “systematically and unfairly discriminate” against individuals or certain social groups, they emphasized that if a biased system becomes widely adopted in society, that the social biases it perpetuates will have serious consequences.
More than 20 years later, the ACM U.S. Public Policy Council (USACM) and the ACM Europe Policy Committee (EUACM) published a joint Statement on Algorithmic Transparency and Accountability,\footnote{\url{https://www.acm.org/binaries/content/assets/public-policy/2017_joint_statement_algorithms.pdf}} underscoring widespread concerns surrounding computer bias, but this time, focusing on the social consequences of \textit{data-driven algorithmic processes and systems}. The statement puts forward seven principles to be considered in the context of system development and deployment, in working toward mitigating the threat of harm to people posed by biases. Despite that the principles are articulated in a single page, it is clear that the issue of algorithmic bias is extremely complex. Multiple sources of bias (e.g., data, modelling processes) are mentioned, as well as alternative solutions -- from simply raising users' awareness of the issue, to enabling the auditing of models by third parties. Furthermore, the principles mention a range of stakeholders (the algorithm's owners, designers, builders, and end users), alluding to their roles in ensuring the ethical development and appropriate use of algorithmic processes. 

Despite the recent surge in attention to the topic, addressing algorithmic bias is not a new concern for researchers. For instance, in the 1990s, machine learning researchers were considering problems of \textit{explainability}, or how to interpret models and facilitate their use (e.g., \cite{craven_using_1994}, \cite{craven_extracting_1996}, \cite{domingos_knowledge_1998}). In the early 2000s, researchers in the data mining community were developing processes for \textit{discrimination discovery} from historical datasets (e.g., \cite{pedreschi_integrating_2009}). Similarly, around the same time, information retrieval researchers were considering the issue of bias in training datasets (e.g., \cite{buckley_bias_2007}) and the resulting impact of this bias on ranking algorithms \cite{cho_impact_2004}. Thus, while several research communities were tackling various issues related to  algorithmic biases earlier on, they were largely disjoint from one another. Furthermore, they addressed the problems from ``inside,'' working exclusively from the perspective of the developer. More recently, multiple perspectives on algorithmic bias have come to light, with the increasing influence of algorithmic systems in society. Arguably, a 2016 article entitled \textit{Machine Bias} \cite{angwin2016machine} played a key role in stimulating widespread discussion, opening up the conversation to other stakeholders beyond those who develop algorithmic processes and systems.

%Integrated solutions for mitigating algorithmic biases are unlikely to be found in one research community alone but rather, must involve work across disciplines. Indeed, this revelation has led to a number of cross-disciplinary initiatives, such as the ACM FAccT Network.\footnote{\url{https://facctconference.org/network/}} FAccT\footnote{Note that in 2020, FAccT became the new acronym of ACM FAT*.} reaches beyond computer science into the social sciences and humanities, as well as law, in addressing fairness, accountability and transparency issues in socio-technical systems. Similarly, many initiatives within artificial intelligence, including workshops and research groups, are focused on FATE (Fairness, Accountability, Transparency and Ethics in AI) (e.g., Microsoft's FATE group\footnote{\url{https://www.microsoft.com/en-us/research/theme/fate/}}).

%In this survey, we aim to facilitate a high-level understanding of the research surrounding the mitigation of bias in algorithmic processes and systems.
Recently, a number of comprehensive surveys has emerged on algorithmic bias, shedding light on the source(s) of bias and highlighting potential solutions. However, such surveys tend to focus on one source of algorithmic bias and/or one class of solutions. For instance, Olteanu and colleagues \cite{olteanu_social_2019} reviewed the literature surrounding data biases; in particular, they address social data sources, given their frequent use in the creation of training data sets. Coming from a fair machine learning perspective, Mehrabi and colleagues \cite{mehrabi2019survey} provided a survey of common problems and solutions, including those focused on data and processes. Addressing explainability, Guidotti et al. contributed a comprehensive survey and a taxonomy of the various methods used to interpret the behaviors of black box models \cite{guidotti_survey_2018}. In addition, there are survey papers providing deep dives into the technical solutions proposed in very well-defined areas. For instance, in \cite{sun2019mitigating}, the authors focus specifically on gender bias in the natural language processing domain, in~\cite{caton2020fairness, chouldechova2018frontiers}, the authors consider the technical approaches of mitigating bias in ML while in~\cite{balayn_managing_2021}, the authors focus on data bias and data management approaches for mitigating bias.

%\green{KO: Here to add the connection of explainability on mitigating bias}

In this survey, our aim is to help the reader achieve a high-level understanding of the current state of this complex topic, across domains. With a view toward promoting more comprehensive solutions, we present a \textit{fish-eye view} of the literature surrounding algorithmic bias, and provide a methodology that is based on three key aspects, namely problems, domains and stakeholders. By examining the literature along these lines, we can better understand how solutions can and should be used to address algorithmic system bias.

In information visualization, fish-eye views, which balance focus and context (i.e., depth and breadth), are useful for facilitating understanding in information spaces that are very large and diverse \cite{furnas2006fisheye}. The user maintains perspective of the ``big picture,'' but can still choose when to drill down into further details. Given the diversity of perspectives 
%\green{and definitions} 
on algorithmic bias, we argue that a high-level view is much needed, particularly for researchers and practitioners new to the area. 

%Thus, we survey related work across four  communities -- machine learning (ML), human-computer interaction (HCI), recommender systems (RecSys), and information retrieval (IR) -- in order to characterize the problems of algorithmic biases that are being addressed, as well as the solutions being proposed, across communities. This allows us to capture perspectives and processes involving multiple stakeholders, as depicted in Fig.~\ref{fig:roles_bias}. For instance, while the ML literature focuses primarily on the developer perspective (and thus, \textit{formal} processes), HCI researchers consider the user's interaction with the system or how the interface might influence the user's perception of fairness (more \textit{informal} processes). IR and RecSys represent communities focused on end user application areas; thus, we can learn the extent to which algorithmic biases have presented challenges to these applications and the nature of the solutions proposed.

%Our goal is to produce a more holistic framework describing the problems of algorithmic bias, as well as the processes and stakeholders involved in addressing them. 
The main contributions of this survey paper are to:
\begin{itemize}
\item Provide a methodology for analyzing the work on algorithmic bias, and a ``live'' repository of articles.% to be enriched over time.
%\item Describe the components of algorithmic processes and systems, which can lead to biases.%, and the dimensions of interest to researchers.
\item Document the problems and solutions studied across research communities.
\item Map the problems to the solutions across diverse domains, as well as the involved stakeholders.
\item Describe opportunities for cross-fertilization between communities, solutions and stakeholders.
%\item Describe how the research communities studied are shaping this cross-disciplinary area.
\end{itemize}

The article is organized as follows. Section~\ref{sec:method} describes the methodology used for the literature review and the selection of domains and publication venues. Section~\ref{sec:approach} presents an overview of the problem and solution spaces discovered while analyzing the papers. Following that, we present the detailed analysis of the three categories of solutions described in the literature: Section~\ref{sec:bias:understanding} focuses on Bias Detection, Section \ref{sec:fairness} details the methods used for Fairness Management, and Section \ref{sec:explain} presents a summary of the work within Explainability Management. In each section, we provide specific examples of the respective solution, described in the literature. Each section ends with a table providing a comparison of the specific approaches taken across the four domains studied. Finally, in Sections \ref{sec:disc} and \ref{sec:open}, we summarize the state-of-the-field, discussing the cross-fertilization among the four communities, the stakeholders and the solutions. These sections also present some open issues for further consideration.

%\section{Definitions}
%\input{./SECTIONS/definitions.tex}

\section{Methodology}~\label{sec:method}
%We first describe the methodology used to identify related articles, as well as the repository we composed. Following that, we describe how we analyzed each article, first identifying the problem(s) addressed in the article (Section \ref{sec:problem}) as well as the type of solution(s) proposed by the authors (Section \ref{sec:sol}).

%\subsection{Methodology for collecting and analyzing articles}
We follow a methodology involving both bottom-up and top-down processes for collecting articles relevant to \textit{bias in algorithmic systems}. The methodology can be characterized as an adaptation of the standard facet-based methodology used in information science to carry out book and even product classification \cite{hjorland2002domain}. In the first phase, a bottom-up, open search process took place, in which each co-author collected relevant literature, adding it to a shared repository. This initial body of material was then used to guide the choice of research domains and publication venues upon which to focus, as well as to identify a set of properties by which to characterize the problems and solutions described. 

\subsection{Selection of Domains}\label{sec:domains}
%Following the development of the guiding concept, and the classification scheme for the problems and solutions, a top-down approach was implemented. 
An inventory of the initial article repository was taken, to understand which domains (i.e., research communities) had produced a critical mass of publications related to the mitigation of algorithmic biases. We focused on well-established domains within the information and computer sciences, which are investigating data and knowledge transformation and communication to the user.
%\green{We did an exhaustive search of the work being carried out in well-established domains within the information and computer sciences, searching for relevant works concerned with data and knowledge transformation and communication to the user.} 
Based on the initial inventory, four domains emerged -- machine learning (ML), human-computer interaction (HCI), recommender systems (RecSys), and information retrieval (IR) -- to characterize the problems of algorithmic biases that are being addressed, as well as the solutions being proposed, across domains. Next, we provide additional justification of these four domains.

The widespread application of ML techniques, which in many cases are opaque, led to the issue of potential bias and discrimination of algorithmic systems and processes. Hence, the ML community and ML-related publications naturally emerged as an established area we needed to review. RecSys represents a specific application area and a domain that attracts significant research attention on algorithmic bias. Within RecSys, ML techniques are applied for reasoning on and exploiting user characteristics; thus, within this domain, many challenges have arisen surrounding potential bias and fairness. IR focuses on information delivery to users, often with the use of search and ranking algorithms that are opaque; thus, bias and fairness have long been researched. The above domains cover a substantial amount of applications where the risk of bias and discrimination in the reasoning process exists. Finally, HCI directly considers the end users and their perceptions when interacting directly or indirectly with different applications. In particular, understanding the potential bias, discrimination or fairness issues that might emerge when a user is interacting with information presented through an interface is considered of high importance. It should be noted that an ``Other'' domain emerged, through the initial repository, where we collected a number of articles published in emerging, cross-disciplinary communities or domains that are not represented in the above main categories. Through ``Other'' we were able to capture research published in other domains where a mass of publications related to bias, fairness and explainability did not (yet) exist, but important work was published, hence, making this a comprehensive review with applicability in areas other than the main domains that emerged. % While these other communities are sensitive to the problems surrounding algorithmic bias, they are much less focused. }

\subsection{Selection of Publication Venues}
Through the exercise of selecting the research domains, a list of high-impact publication venues, including both conferences and journals, was created for each domain, as presented in Table~\ref{tab:pub:domain}. Also, note that some venues publish articles across domains. For instance, while ACM CSCW is generally aligned with the HCI community, some articles describing studies of recommender systems can be found there. Such cases are indicated with parentheses in Table~\ref{tab:pub:domain}.

\begin{table}[ht]
    \centering
    \begin{tabular}{|c|c|c|}
    \hline
         Domain & Publication Venues Reviewed & \# Papers  \\
         \hline
         Machine Learning/AI & AAAI, IJCAI, KDD, SIGKDD, CIDM, ICML, AIES, NIPS, & 106\\
         &MLSP, ACM Data Mining and Knowledge Discovery Journal & \\ \hline
         %&IJCAI&\\& KDD &\\& SIGKDD &\\ & CIDM&\\& AIES&\\& NIPS & \\&MLSP &\\ & ACM Data Mining and Knowledge Discovery Journal & \\\hline
         Information Retrieval & ACM SIGIR, ACM CIKM, ACM WWW, & 68\\
         &TOIS, JASIS, IR Journal, (AAAI ICWSM) & \\ \hline
         %& ACM CIKM &\\ & ACM SIGIR&\\& ACM WWW &\\& TOIS &\\ & JASIST &\\& IR Journal & \\\hline
         Recommender Systems & ACM RecSys, AAAI ICWSM, UMUAI, ArXive & 46 \\  
         & (ACM CSCW, ACM CIKM, ACM FAccT (formerly FAT*)) & \\ \hline
         % & UMUAI, ArXive & \\ \hline
         %& ACM WWW CHI CSCW &\\& ACM RecSys &\\ & ArXive &\\& ACM FAT* &\\& UMUAI &\\  & ACM SIGIR & \\\hline
         Human Computer Interaction & ACM CHI, ACM CSCW, ACM CHI Journal, CSCW Journal &  34\\
         & Journal of Behaviour and Information Technology & \\ 
         %& ACM CHI &\\& CSCW Journal&\\& ACM HCI Journal&\\& INTERACT &\\& Journal of Behaviour and Information Technology &\\& Journal of Big Data and Society & \\
         \hline
         Other & AAAI HCOMP, ACM FAccT, ICDM,VLDB & 57\\
         %& ACM FAT* &\\
         \hline

    \end{tabular}
    \caption{Key publication venues reviewed per domain.}
    \label{tab:pub:domain}
\end{table}

The next step was to review each publication venue’s proceedings / published volumes during the twelve-year period 2008 - 2021, resulting in a high-recall search for relevant published articles. The key words used were: ``accountability,” ``bias,” ``discrimination,” ``fair(ness),” ``explain(able),” and ``transparen(cy).” In addition, the articles collected address a particular algorithmic process or system, or a class of system. In other words, articles of a more abstract or philosophical nature were excluded from the survey. Likewise, in the ML category, articles from AI venues (e.g., AAAI, IJCAI) that were not published in the respective ML track, have been excluded, as to focus on algorithmic, data-driven systems.

\section{Analysis of Papers}~\label{sec:approach}
 This survey is based on our current repository of over 245 articles.\footnote{Available at Zotero - \url{https://www.zotero.org/groups/2450986/cycat_survey_collection_public}.} The list of publication venues reviewed is not exhaustive; further venues may be added to our repository in the future. However, the problem and solution spaces discovered and detailed below, have proven to be robust across the articles reviewed to date. In our repository, each article is labeled with its respective domain (ML, HCI, RecSys, IR, Other). %It is natural that some publications are interdisciplinary (e.g., a study of a recommender system's interaction with a user, which would appear to relate to both RecSys and HCI); thus, the classifications were performed with respect to the publication venue. 
After reviewing the article, three additional properties, which shall be explained below, were also recorded:
\begin{itemize}
    \item The problem(s) identified within the system.
        \item The attribute(s) affected by the problem(s).
    \item The solution(s) proposed to address the problem(s).
\end{itemize}

These four attributes -- domain, problematic system component(s), attribute(s) affected by the problem, and proposed solution(s) -- are provided as tags in our Zotero repository. Thus, other researchers may use this resource in various ways, e.g., to focus on a specific problem or type of solution. Table~\ref{tab:annotation-examples} provides examples of the manner in which articles in our repository were analyzed; further details are provided in the following subsections.

\begin{table}[ht]
    \centering
    \footnotesize 
    \begin{tabular}{|c|c|c|c|c|}
    \hline
    Domain & Example & Problem(s) & Attribute(s) & Solution(s) \\ \hline
 ML&Word embeddings trained on Google News &Data &gender &Discrimination \\ 
 & articles were found to perpetuate prevalent   & &  & discovery - indirect \\ 
 & gender biases. \cite{bolukbasi_man_2016} & & & \\ \hline 
IR &Users of Mendeley search were shown to  & User &national origin & Discrimination \\ 
   & disproportionately favor articles written by & &                 & discovery - direct \\ 
   &authors sharing their national origin. \cite{thelwall_are_2015} & & & \\ \hline 
   RecSys &Profiles of women and people of color in &  User, Data&gender, race & Auditing \\ 
          &online freelance marketplaces were found & Third parties & & \\
          &to be systematically lower-ranked than others; & & & \\
          &reasons included bias in training data and  & & &\\
          &lower evaluations by other users. \cite{hannak_bias_2017} & & & \\ \hline
HCI &Authors provided various explanations to& User&information &Explainability - \\
    & users about their Facebook feeds. & Output & & model, output \\ 
    &Explanations were found to shape beliefs on & & & \\ 
    &how the system works, but not in   & & & \\ 
    &understanding its specific outputs. \cite{rader_explanations_2018} & & & \\ \hline 
Other    &Authors addressed the issue of human bias in  &Data &information & Fairness pre-processing\\
         &computer vision training data, through an & & &  \\ 
         &algorithm that filters human reporting bias & & & \\
         &from labels that are visually grounded. \cite{misra_seeing_2016} & & & \\ \hline
    \end{tabular}
    \caption{Example analyses of articles in the repository.}
    \label{tab:annotation-examples}
\end{table}

\subsection{Problem space} \label{sec:problem}
To explore the problem space within the literature addressing algorithmic bias, we characterized, for each article, the system component(s) deemed to be problematic by the authors, as well as the attribute(s) whose values are affected by the bias. 
%The definition of the problem state space of an algorithmic system consists of detecting the bias in the system and identifying the source/s of the bias using different approaches. Bias in algorithmic systems is generated mainly by humans through the collected data and the algorithms may reproduce or increase existing discrimination bias from the input data. Moreover, bias can be caused by expectations (third parties or developers). 
\subsubsection{Problematic system component(s)}
\begin{figure}[ht]
    \centering
    \includegraphics[width=0.7\textwidth]{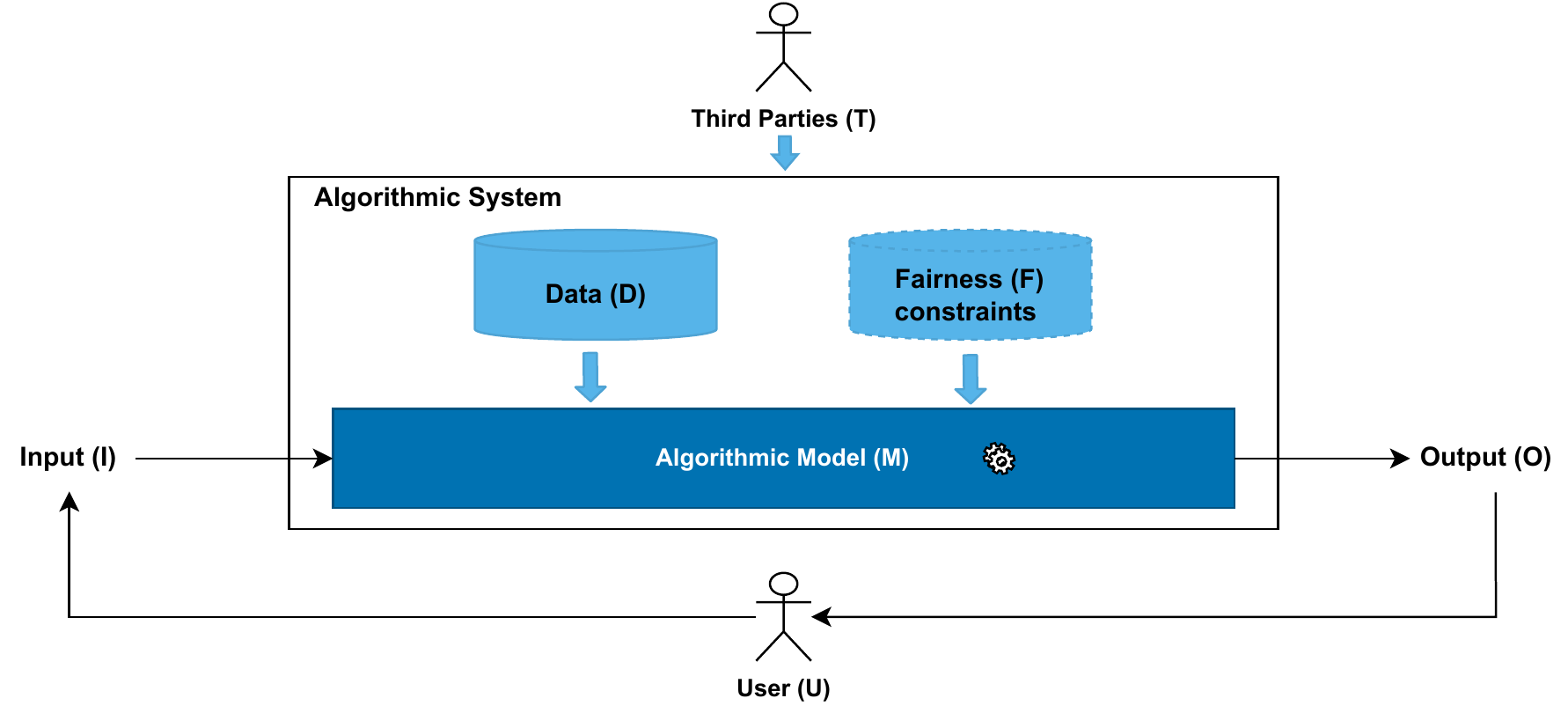}
    \caption{Generic architecture of an algorithmic system.}
    \label{fig:alg_system}
\end{figure}

We recorded the macro component(s) of the algorithmic system or process,\footnote{Henceforth, we refer to a ``system'', although as mentioned, we consider articles that describe particular algorithmic processes as well as those describing deployed systems.} considered by the author(s) as being the source of the problem. Figure~\ref{fig:alg_system} provides a general characterization of an algorithmic system, with its macro components, which we have used to examine the problem space of algorithmic bias. Note that some components are optionally present. This includes a User (U), who interacts directly with the system's inputs and/or outputs. Alternatively, an API may be in place, to allow the system to interact with other systems and applications.

In this generic architecture, the system receives input (I) for an instance of its operation. This is provided by a user (U), or another source (e.g., the result of an automated process). The algorithmic model (M) makes some computation(s) based on the inputs and produces an output (O). The model learns from a set of observations of data (D) from the problem domain. It may also receive constraints from third-party actors (T) and/or internal fairness criteria (F) that modify the operation of the algorithmic model (M). Finally, some systems have direct interaction with a user (U) who, as previously discussed, will bring her own knowledge, background and attitude when interpreting the system's output.

Thus, as depicted in  Figure~\ref{fig:alg_system}, bias may manifest and/or be detected in one or more of these components:
\begin{itemize}
\item Input (I) - Bias may be introduced in the input data, e.g., as incorrect or incomplete information input by the user.
\item Output (O) - Bias may be detected at the outcome (value(s)/label(s)) produced in response to the input. %e.g., incorrect labels or at the interpretation of the decision outcome.
\item Algorithm (M) - Bias can manifest during the model's processing and learning.
\item Training Data (D) - Training data may be unbalanced and/or discriminatory toward groups of people. The data may also be based on an unrepresentative set of instances, and may also suffer from inaccuracies in the ground truth. 
%Information about sensitive attributes of people may be contained in the training data and such information may be unbalanced and/or discriminatory toward groups of people. The data may also be based on an unrepresentative set of instances, and may also suffer from inaccuracies in ``ground truth." 
\item Third Party Constraints (T) - Implicit and explicit constraints, given by third parties, may impact the design and performance of the algorithm such as to be discriminatory towards groups of people. These include operators of the system, regulators, and other bodies that influence the use and outcomes of the system.\footnote{An example was described in Table~\ref{tab:annotation-examples} of a RecSys in which other users' ratings of workers affected system performance during a given user's instance. Another example might be a search engine suppressing some ranked results to comply with laws in the user's geographical region.}
%\item Fairness Constraints (F) - Criteria introduced within the system that can change algorithmic , such that one interpretation of fairness is prioritized over others.
\item User (U) - When users interact directly with a system, they may contribute to bias in a number of ways, such as through the inappropriate use of the system or misinterpretation of system output. 
\end{itemize}

%copy from Deliverable D4.1 - needs rephrasing

%Further to the above, two additional sources of bias were encountered in the literature. First, as previously mentioned, there are often competing fairness constraints, which cannot be simultaneously addressed \cite{kleinberg_inherent_2016}. Since addressing one often means that the other will be left unaddressed, the fairness constraints themselves may be the source of bias (F). 

%The user's (U) perception can serve an additional source of bias. As mentioned previously, the manner in which a user interprets the system's behaviors does not only depend on the output, but is also related to her own perception, background, and attitude. 
%Thus, perceived bias can be caused or get influenced by the input or output of the algorithmic system. It can be related to the user's input knowledge that can influence the training data and may create an algorithm that is perceived to be \lq{}unfair\rq{}.

\subsubsection{Affected attribute(s)}\label{sec:attr}
We have also characterized, for a given article, the attribute(s) whose values are influenced by the problematic system behavior. While early technical papers used the generic terms, \textit{``sensitive'' or ``protected'' attributes}, to characterize the features on which a group is unfairly treated by the algorithmic decision~\cite{pedreschi_measuring_2009}, recent work has considered a broader set of attributes, including the social, cultural, and political attributes of the content or person being processed,  e.g., gender, race, age, income, etc.  

The system under study in each article can exhibit different behaviors with respect to the 
affected attribute, which may or may not be problematic for a given user or observer. It can be noted that while many of the affected attributes concern social and cultural characteristics (i.e., characteristics describing the social world), we also observe dimensions such as the quality / accuracy / credibility of the information provided to the user (i.e., information attribute). By information, we mean the quality (or bias) of the information that is conveyed by or to the user. In other words, the concern here is the extent to which the data used by the algorithm constitute a truthful representation of the world. Note that information bias may also be introduced by the algorithm because of its low classification/predictive performance, i.e., low accuracy. Even though such instances may not represent cases where an algorithmic system’s behavior can directly result in discrimination or harm, in many contexts, these issues can indirectly lead to serious consequences for system users (e.g., limited exposure to high-quality sources of information on a given topic because of biased search engine results).

Information is the most studied attribute in our corpus, and is the primary dimension addressed in the ML literature. For instance, in the explainability literature, a primary concern is the extent to which information is effectively conveyed to the user. Likewise, IR articles often consider information as the affected dimension under study; here, the classic example is the large body of work on search engine biases. In contrast, the literature in HCI and RecSys does not often address information as an affected dimension. In these fields, articles on mitigating algorithmic biases more often consider social and cultural dimensions, such as demographics as a general term, or more specific attributes such as gender and race, with a few studies emerging on characteristics such as age, language and physical attractiveness. 

\subsection{Solution Space}\label{sec:sol}
The solution space discussed in the literature we surveyed consists of three main steps in mitigating algorithmic bias. Each of these steps involves different stakeholders within each community. Next, we describe the multiple stakeholders who are involved in the solution processes. Afterwards, we give a detailed overview of each of the three steps in bias mitigation. 

\subsubsection{Stakeholders}
\begin{figure}[ht]
    \centering
    \includegraphics[width=0.8\textwidth]{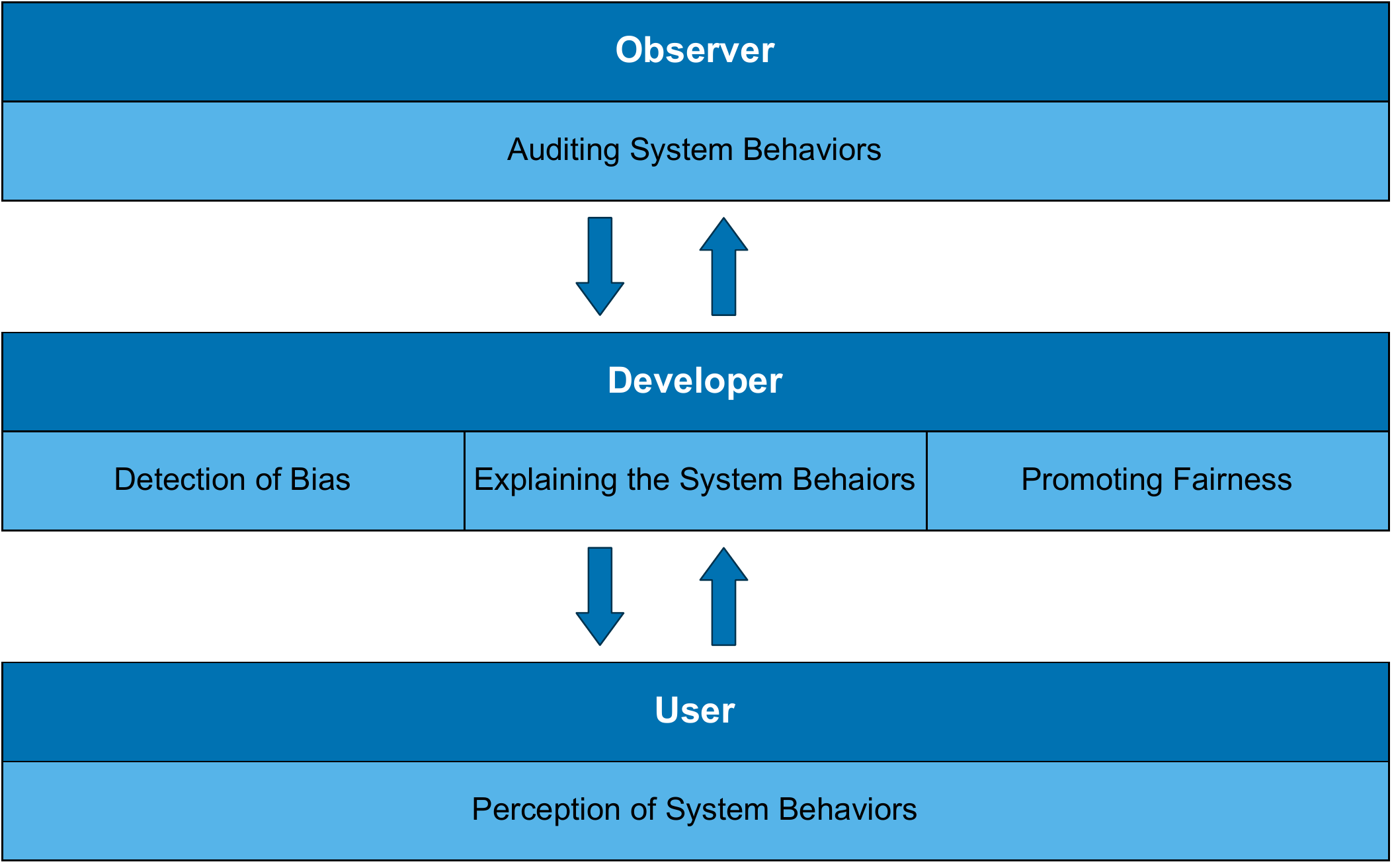}
    \caption{Processes and stakeholders involved in mitigating algorithmic bias.}
    \label{fig:roles_bias}
\end{figure}

The selection of the four domains allows us to capture perspectives and processes involving multiple stakeholders, as also depicted in Figure~\ref{fig:roles_bias}. For instance, while the ML literature focuses primarily on the developer perspective (and thus, \textit{formal} processes), HCI researchers consider the user's interaction with the system, or how the interface might influence the user's perception of fairness (and thus, more \textit{informal} processes). IR and RecSys represent communities focused on end user applications; thus, we can learn the extent to which algorithmic biases have presented challenges to these applications and the nature of the solutions proposed.

\textit{Developer(s)} can internally detect bias in data and processes, evaluate formal notions of fair treatment of the individuals, groups and/or content affected by algorithmic judgements, 
%for whom recidivism scores are predicted, 
as well as implement methods used by the system for explaining its decisions to users and/or third parties. \textit{System Observer(s)}, who may be regulators, researchers or even data journalists, can conduct their own audits of the system behaviors. However, \textit{User(s)} of the system have their own perceptions of the system's behavior, which depend not only on the system itself, but also their own knowledge, experience and attitudes.
\textit{Indirect User(s)} are the people who are affected by the algorithmic decision. These are, for instance, defendants evaluated by an algorithmic risk-assessment system or candidates whose resumes are screened with an algorithmic resume screening system. Indirect users also have their own perceptions of the system’s behavior.

%\green{KO: Give definitions of information and explain general terms demographics, sensitive attributes, protected groups}

\subsubsection{Classification of Solutions}
The literature suggests that a comprehensive solution for mitigating algorithmic system bias consists of three main steps: $i$) Bias Detection, $ii$) Fairness Management, and $iii$) Explainability Management.
%\begin{itemize}
%\item Detection of Bias
%\item Fairness Management 
%\item Explainability Management 
 %   \item \textbf{Discrimination discovery} which includes the search through the algorithmic system to detect any type of discrimination.
 %   \item \textbf{Fairness management and certification} approaches are then applied to mitigate the detected bias and certify that the system is fairness-aware.
 %   \item \textbf{Explainability management} approaches have to be applied to the system to force transparency and to build trust between the end user and the system.
 
\textbf{Bias Detection} includes techniques that scrutinize the system to detect any type of systematic bias. It can be achieved by \textit{Auditing} and/or \textit{Discrimination Discovery} methods.  Auditing involves making cross-system or within-system comparisons, and is typically done by an analyst / observer or a regulator who does not have access to the inner-workings of the system~\cite{sandvig_auditing_2014}. There are variations in the extent to which the auditing approaches are formalized. In some cases, auditing uses the tools of discrimination discovery (e.g., discrimination/fairness metrics). In this sense, auditing as a term refers to who is doing the discrimination discovery and why, but not to a different set of tools and techniques. In other cases, auditing is used in a more formal way to detect any fairness issues in the system. Discrimination discovery (direct or indirect) approaches include tools and practices for detecting unfair treatment by data / algorithms / systems using statistical metrics, i.e., measuring specific fairness notions.
%apply to the study of particular tasks and/or algorithms (e.g., a top-k ranking algorithm) as well as to deployed systems, which may consist of a whole collection of algorithmic processes (e.g., a proprietary search engine, which uses not only relevance ranking, but also personalization / localization algorithms, among others). In sum, discrimination discovery 
%Bias detection is reviewed in detail in Section~\ref{sec:bias:understanding}.
 %Multiple discrimination refers to cases where an individual is discriminated against on the basis of two or more characteristics, sometimes referred to as \textit{sensitive attributes}~\cite{romei_multidisciplinary_2014}.

%The underlying cause of discrimination is typically unintentional prejudice in an individual or group, or when economic agents (e.g., consumers, workers, employers) lack adequate information about individuals with which they interact (i.e., statistical thinking). 
\textbf{Fairness Management} includes techniques that developers use to mitigate the detected bias and certify that the system is fairness-aware. Fairness management approaches are used by developers to tackle bias in different parts of the system. They are divided into: \textit{Fairness pre-processing}, \textit{Fairness in-processing}, \textit{Fairness post-processing}, \textit{Fairness certification} and \textit{Fairness perception}.

\begin{itemize}
%\item \textit{Auditing (Fairness formalization)} where the fairness constraints or any fairness measures, specifications and criteria for the system design are defined. Different ways of addressing fairness in the system are also defined.
\item \textit{Fairness Pre-processing} includes approaches that process the training data in a manner that promotes fairness. Examples are: re-sampling, re-weighting and feature transformation approaches.

\item \textit{Fairness In-processing} includes approaches that address discrimination during the training procedure. Examples are: regularization, optimization and learn-to-rank approaches. %In user-focused systems, fairness learning solutions impose constraints that force the learner to result in fairer models.
\item \textit{Fairness Post-processing} includes approaches that ensure that a system is ``fair'' by changing the output of the learned classifier i.e., changing the label weights and re-ranking approaches. 
\item \textit{Fairness Certification} includes approaches provided by the developer/observer in the case where no unintended bias has been detected in the system. The developer/observer verifies whether the output satisfies specific fairness constraints, and if so, he or she can certify the algorithmic system as ``fair." In general, fairness certification aims to test algorithmic models for possible disparate impact, according to the fairness internal certification and bias detection results, ``certifying" those that do not exhibit evidence of unfairness.
\item \textit{Fairness Perception} includes approaches that examine the perception of different stakeholders with the decision outcome of the algorithm. Examples are the use of questionnaires and statistical tests. 
\end{itemize}
%is a method used to (internally) certify that a system is fairness-aware \green{based on fairness metrics and constraints. \textit{Fairness perceived management concerns the perception of users with the decision making outcome and it can be measured through questionnaires and statistical tests ~\cite{lee_algorithmic_2017}. According to~\cite{lee_algorithmic_2017}, the perceived fairness of decisions depends on the task and whether the decision maker is a human or an algorithm. For mechanical tasks, user perception on fairness and trust of the system is similar, whether the decision is taken by a human or an algorithm. However, in tasks of a more social nature, a user perceives the human decision making outcome as more fair and trustworthy rather than a decision outcome made by an algorithm.
 %Section~\ref{sec:fairness} reviews the techniques described in the literature for fairness management. 

\textbf{Explainability Management} includes techniques that facilitate transparency and build trust between the end user and the system.  Explainability and interpretability contribute to the sense of transparency as well as the perception of fairness~\cite{explainability2021}. Explainability approaches are used to provide transparency of the system and in that way, enable the detection of any bias or fairness issues in the data and model. In general, explainability-aware techniques can be divided into two main categories:
\textit{Model Explainability}, which provides explanations for the training process of the models and \textit{Outcome Explainability}, which provides explanations of the algorithm's decision outcome in an understandable way to the user. Outcome Explainability methods explain only the output, and they do not provide explanations for the process of the algorithm. This form of explanation is usually helpful when the user of the system is not an expert, such as in the case of RecSys. %The literature on Explainabiity is presented in Section~\ref{sec:explain}.

\begin{figure}[ht]
    \centering
    \includegraphics[width =0.9\textwidth]{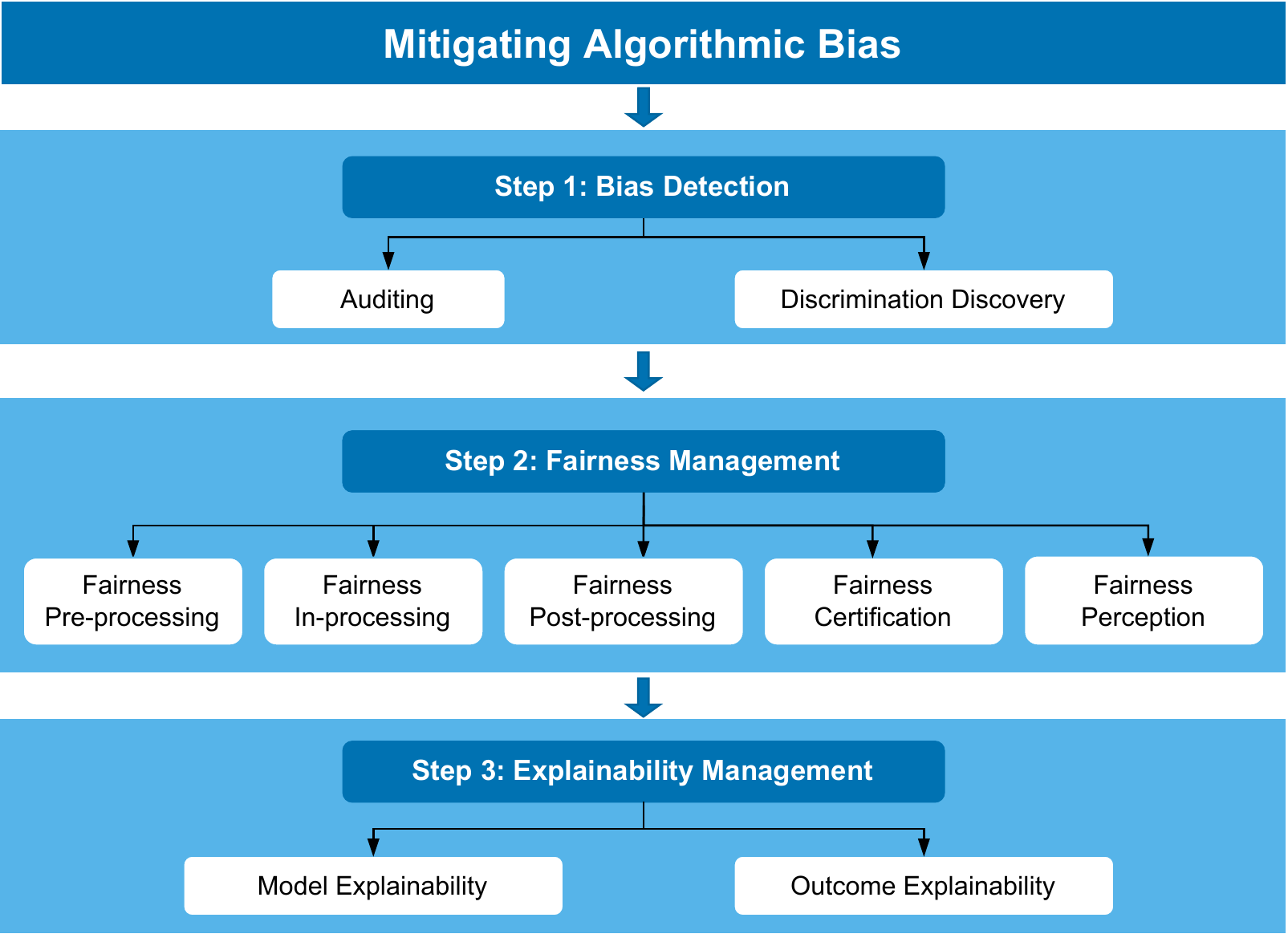}
    \caption{The solution space - tools for mitigating bias in algorithmic systems.}
    \label{fig:FAT:solutions}

\end{figure}

Figure~\ref{fig:FAT:solutions} aligns the three steps involved in mitigating biases, with the taxonomy of solutions found in the literature surveyed.  %In ML and AI systems, these approaches are divided into white-box and black-box explainability methods regarding the interpretability of the algorithm that is used in the system. In the user-focused systems, explainability aims to provide transparency to the user for the outcome decision of the system.

%\green{KO: Should we include the disclaimer about our classification here??}
\subsection{Summary}
Before describing each set of techniques in detail, we provide a summary overview of the problems and solutions documented within each of the four domains surveyed. These are presented for each of the three steps (i.e., Table~\ref{tab:sol:bias} presents Bias Detection solutions, while Tables~\ref{tab:sol:fairness} and ~\ref{tab:sol:explain} present solutions for Fairness and Explainability Management, respectively). The distribution of problems addressed across the four domains illustrates the insights gained from our ``fish-eye view." As expected, the ML literature addresses problems concerning the training data, the algorithmic model and the system output. The RecSys and IR literature, as user-focused application areas, consider problems both inside and outside the system, while HCI naturally addresses the interactions between the user and the algorithmic system. 

Similarly, we find that across domains, researchers are engaged in all three steps in bias mitigation -- detection, fairness and explainability management. In the following sections, we shall provide a detailed overview of specific examples of the approaches of each of the three steps, from across domains, and shall compare the techniques used. 

\begin{table}[ht]
\small
    \centering
    \begin{tabular}{|l|l|l|l|}
    \hline
         \textbf{Domain} & \textbf{Problem} & \textbf{Solution} & \textbf{Reference(s)} \\
         \hline
         \multicolumn{4}{|c|}{\textbf{Bias Detection}}\\\hline
         ML & Data/Model &Auditing & \cite{Luong2011, zhang_causal_2017, saleiro_aequitas:_2018, asplund_auditing_2020} \\
        % \hline
          & Data/Model/Output & Discrimination Discovery &  \cite{zliobaite_survey_2015, zafar:2017, leavy_gender_2018, zhang_causal_2017, pedreschi_measuring_2009, datta_algorithmic_2016, cowgill_algorithmic_2017, jiang_reasoning_2020}\\
         \hline
         IR & User/Data & Auditing & \cite{vincent_measuring_2019, kay_unequal_2015, magno_stereotypes_2016, otterbacher_competent_2017, kulshrestha_quantifying_2017, hu_auditing_2019, le_measuring_2019}\\ 
         
         & User/Data & Discrimination Discovery &  \cite{weber_demographics_2010, yom-tov_demographic_2019, pal_exploring_2012, chen_investigating_2018, wilkie_retrievability_2014,wilkie_best_2014, wilkie_algorithmic_2017,bashir_relationship_2011,eickhoff_cognitive_2018, lin_impact_2015, kodama_theresa_2017}\\
         %&%User/Output& &\cite{jansen_examination_2006, barilan_presentation_2009, white_belief_2014, white_belief_2015,  }\\
          
          \hline
          HCI & User/Data & Auditing& \cite{matsangidou_what_2019, johnson_effect_2017, metaxa_image_2021}\\
          & Third Party/Model/Output & Discrimination Discovery&  \cite{das_gendered_2019,quattrone_theres_2015, green_disparate_2019, barlas_what_2019}\\ \hline
          RecSys &Data/Output& Auditing &  \cite{eslami_be_2017, edelman_racial_2017, imana_auditing_2021}\\
         &Data/Output& Discrimination Discovery & \cite{ali_discrimination_2019, speicher_potential_2018, sweeney_discrimination_2013, bellogin_statistical_2017, ekstrand_all_2018}\\
        \hline
        \end{tabular}
        \caption{Summary of the problem and bias detection solution space per domain.}
         \label{tab:sol:bias}
        \end{table}
        
        \begin{table}[ht]
        \small
            \centering
            \begin{tabular}{|l|l|l|l|}
            \hline
            \textbf{Domain} & \textbf{Problem} & \textbf{Solution} & \textbf{Reference(s)} \\
         \hline
        \multicolumn{4}{|c|}{\textbf{Fairness Management}}\\\hline
        ML & Data & Fairness Pre-processing & \cite{calders2010three, johndrow_algorithm_2019,zhang_causal_2017, l._cardoso_framework_2019, kamiran2009classifying}
        \\
       % \hline
        & Model &Fairness In-processing & \cite{zhang_fairness_2018, celis_classification_2019, kleinberg_inherent_2016, dimitrakakis_bayesian_2018,khademi_fairness_2019, grgic-hlaca_human_2018, wu_convexity_2019,kamishima_fairness-aware_2012,kusner_counterfactual_2017,kilbertus_blind_2018, yan_fairness-aware_2020, rezaei_fairness_2020}\\
       % \hline
         & Model/Output &Fairness Post-processing & \cite{kamiran2009classifying, hardt_equality_2016, pedreschi_measuring_2009} \\
        &User/Output& Fairness Perception& \cite{srivastana:2019}\\
        &Data/Model/Output &Fairness Certification &\cite{fang_achieving_2020, kilbertus_blind_2018, sharma_certifai_2020, zhang_fairness_2018, celis_classification_2019, kleinberg_inherent_2016, dimitrakakis_bayesian_2018, khademi_fairness_2019, grgic-hlaca_human_2018, wu_convexity_2019, cruz_cortes_invitation_2020}\\
       \hline
       
        IR & Data &Fairness Pre-processing & \cite{gjoka_walking_2010, dixon_measuring_2018, shen_darling_2018, diaz_addressing_2018}\\ 
        %\hline
        & User/Model &Fairness In-processing &\cite{dai_adversarial_2021, ovaisi_correcting_2020, yang_maximizing_2021}\\ 
        & User/Output &Fairness Post-processing & \cite{karako_using_2018, kirnap_estimation_2021, kuhlman_fare:_2019}\\
         & User & Fairness Certification&\cite{epstein_suppressing_2017, hofmann_eye-tracking_2014, mitra_user_2014}\\
        \hline
        HCI &  Data& Fairness Pre-processing methods& \cite{johnson_effect_2017}
         \\
        & User/Model &Fairness Perception& \cite{brown_toward_2019}\\
        & User/Output &Fairness Certification &\cite{woodruff_qualitative_2018, lee_algorithmic_2017}\\
      
        \hline
        RecSys & Data & Fairness  Pre-processing methods& \cite{burke2018balanced, wu_learning_2021,Luong2011, karako_using_2018, xiao_fairness-aware_2017, mehrotra_towards_2018, yang_maximizing_2021}\\
        %\hline
        & User/Model &Fairness In-processing methods & \cite{yang_maximizing_2021}\\
         & User/Output &Fairness Post-processing & \cite{karako_using_2018, zehlike_fair:_2017, singh_fairness_2018}\\
        \hline
       
       \end{tabular}
      \caption{Summary of the problem and fairness management solution space per domain.}
       \label{tab:sol:fairness}
       \end{table}
      
       \begin{table}[ht]
       \small
           \centering
           \begin{tabular}{|l|l|l|l|}
         \hline
         \textbf{Domain} & \textbf{Problem} & \textbf{Solution} & \textbf{Reference(s)} \\
         \hline
       \multicolumn{4}{|c|}{\textbf{Explainability Management}}\\\hline
       ML & Model&  Model Explainability &\cite{krishnan_extracting_1999, boz_extracting_2002, domingos_knowledge_1998, gibbons_cad-mdd:_2013, chipman_making_2007, zhou_learning_2016, schetinin_confident_2007, tan_tree_2016} \\&&&\cite{ craven_extracting_1996, johansson_evolving_2009, zhou_extracting_2003, tan_detecting_2017, lu_explanatory_2006, card_deep_2019}\\
       
        & Output & Outcome Explainability& \cite{strumbelj_efficient_2010, ribeiro_why_2016, datta_algorithmic_2016, poulin_visual_2006, ribeiro_why_2016,  ribeiro_anchors:_2018, turner_model_2016}\\&&&\cite{ bojarski_visualbackprop:_2016, zintgraf_visualizing_2017, xu_show_2015, fong_interpretable_2017, selvaraju_grad-cam:_2017, zhou_learning_2016, simonyan_deep_2013,henelius_peek_2014, vidovic_feature_2016}\\
      \hline
       HCI & User & Model Explainability & \cite{horne_rating_2019}\\
       %\hline
        & User/Output & Outcome Explainability& \cite{friedrich2011taxonomy, rader_explanations_2018, eiband_bringing_2018, binnis_its_2018}\\
       \hline
       RecSys & User/Output & Outcome Explainability&\cite{nunes_systematic_2017,wang_explainable_2018, tintarev2007, kouki_personalized_2019, verbert2013visualizing, bountouridis_siren:_2019,cheng_mmalfm:_2019} \\
       \hline
       \end{tabular}
    \caption{Summary of the problem and explainability management solution space per domain.}
    \label{tab:sol:explain}
         \end{table}
         
\newpage

\section{Detection of Bias in Algorithmic Systems}\label{sec:bias:understanding}
%discrimination_detection.tex
There are multiple notions of fairness that are important in the context of an algorithmic system as given in~\cite{verma:2018}. The fairness of an algorithmic model (or classifier), depends on the notion of fairness one wants to adopt. Based on~\cite{verma:2018}, there are three main categories for fairness notions: $i$) Statistical measures, $ii$) Similarity-based measures, and $iii$) Causal reasoning. Before going into a detailed overview of examples for bias detection approaches, we provide the definition of the most popular fairness metrics used in the approaches (discrimination discovery and fairness management) proposed in our corpus.

%\subsection{Fairness Definitions}\label{sec:defini}
 %(Sections~\ref{sec:bias:understanding} - \ref{sec:explain})}.

\begin{itemize}
\item \textit{Demographic parity (or Statistical parity)}~\cite{zliobaite_survey_2015}: Both subjects of the protected and unprotected group have equal probability to be assigned to the positive predicted outcome.
%\item \textit{Disparate impact}~\cite{feldman_certifying_2015}: where the difference of the probability distributions of the outcomes conditional on the predicted variable do not exceed a specific threshold, 
\item \textit{Equality of opportunity (or False negative error balance)}~\cite{hardt_equality_2016}: A statistical group fairness notion. The model satisfies this definition if both subjects of protected and unprotected groups have equal false negative rate (FNR), the probability of an individual who is actually in a positive class to be assigned by the classifier a negative predictive value.
\item \textit{Disparate mistreatment (or Equalized odds)}~\cite{zafar:2017, hardt_equality_2016}: A statistical group fairness notion. The model satisfies this definition of fairness if subjects of both protected and unprotected groups have equal true positive rate (TPR) and false positive rate (FPR).
\item \textit{Counterfactual fairness}~\cite{kusner_counterfactual_2017}: A causal reasoning, individual fairness notion. The algorithmic model is considered as fair when the prediction of an individual is the same even if the value of the sensitive variable changes. To validate this type of fairness, a causal model is used.
\end{itemize}

In addition to the above fairness notions, in ranking systems, such as RecSys and IR, a common type of bias is the \textit{position bias} where users tend to consider only the items ranked in the top few positions~\cite{pitoura_fairness_2021}. The fairness notions that consider position bias are the producer or item-side fairness and the consumer or user-side fairness. Producer or item-side fairness focuses on the items that have been recommended so that similar items are ranked/recommended in a similar way. Consumer or user-side fairness focuses on the users who receive the ranking results or the recommendations. A similar group of users should all receive similar recommendations.

%Here: to include fairness notions for RecSys

%As mentioned in Section~\ref{sec:approach}, the main approaches described in the literature for understanding and detecting \green{bias, discrimination and fairness issues} in an algorithmic system are classified into two categories: \textit{Auditing} and \textit{Discrimination Discovery}. 
Next, we present the details described in the papers collected from the four communities for detecting any type of bias in a system using Auditing and Discrimination Discovery approaches.
%\green{Bias resulting in social discrimination is often the result of training data that encodes sensitive features concerning people and their causal influences. This type of bias can also result from the imbalanced representation of the protected groups in the training data. Because training data come in different modalities and representations such as text data, images, big data, video, etc., the study of discrimination detection in data has raised the interest of various research communities.}
%\subsection{Bias Detection Approaches}
 %\textit{Auditing} approaches include the papers that... while \textit{Discrimination discovery} include the papers that detect bias through a more formal methodology i.e. computing statistical metrics.}

\subsection{Auditing Approaches for Bias Detection}
%Auditing approaches that are used on any algorithmic system for detecting bias through the data are described in ~\cite{romei_multidisciplinary_2014}. The simplest case is when auditors (testers) \green{evaluate the system to detect any bias (input, algorithmic or output)} based on specific criteria or for a given group of individuals. Special cases of auditing approaches are situational and corresponding testing approaches. 

The most common auditing approach used for bias detection involves humans (external testers, researchers, journalists or the end users) acting as the auditors of the system. In information retrieval systems, researchers usually perform an audit by submitting queries to search engine(s) and analyzing the results. For instance, Vincent et. al.~\cite{vincent_measuring_2019} performed an audit on Google result pages, where six types of important queries (e.g., trending, expensive advertising) were analyzed. The goal was to examine the importance of user-generated content (UGC) on the Web, in terms of the quality of information that the search engines provide to users (i.e., if there was a bias in  favor/penalizing such content). Similarly, Kay et. al.~\cite{kay_unequal_2015}, Magno et. al.~\cite{magno_stereotypes_2016}, and  Otterbacher et. al.~\cite{otterbacher_competent_2017} submit queries to image search engines to study the perpetuation of gender stereotypes, while Metaxa et al.~\cite{metaxa_image_2021} consider the impact of gender and racial representation in image search results for common occupations. They compare gender and racial composition of occupations to that reflected in image search and find evidence of deviations on both dimensions. They also compare the gender representation data with that collected earlier by Kay et al.~\cite{kay_unequal_2015}, finding little improvement over time. 

Another example of bias detection in a search engine via auditing is the work of Kilman-Silver et. al.~\cite{kliman-silver_location_2015} who examine the influence of geolocation on Web search (Google) personalization. They collect and analyze Google results for 240 queries over 30 days from 59 different GPS coordinates, looking for systematic differences. In addition, Robertson et. al.~\cite{robertson_auditing_2018} audited Google search engine result pages (SERPs) collected by study participants for evidence of filter bubble effects. Participants in the study completed a questionnaire on their political leaning and used a browser extension allowing the researchers to collect their SERPs. 
  
Kulshrestha et al.~\cite{kulshrestha_quantifying_2017} propose an auditing technique where queries are submitted on Twitter, to measure bias on Twitter results as compared to search engines. The proposed technique considers both the input and output bias. Input bias allows the researchers to understand what a user would see if shown a set of random items relevant to her query. The output bias isolates the bias of the ranking mechanism. In addition, Johnson et. al.~\cite{johnson_effect_2017} investigate the demographic bias detection in Twitter results using as an auditing technique, the retrieval of geotagged content using Twitter API. Another example where researchers are the auditors is the study of Edelman et. al.~\cite{edelman_racial_2017} where the authors run an experiment to audit Airbnb applications to detect racial bias in ranked results, and more specifically, for African American names.

Another cluster of user-based studies in IR systems concerned the detection of perceived biases about search and/or during a search for information. In these studies, users are the auditors. For instance, Kodama et al.~\cite{kodama_theresa_2017} assessed young people’s mental models of the Google search engine, through a drawing task. Many informants anthropomorphized Google, and a few focused on inferring its internal workings. The authors called for a better understanding of young people’s conceptions of search tools, so as to better design information literacy interventions and programs. In addition, Otterbacher et al.~\cite{otterbacher_investigating_2018} described a study in which participants were the auditors for detecting perceived bias. They were shown image search results for queries on personal traits (e.g., “sensitive person”, “intelligent person”) and were asked to evaluate the results on a number of aspects, including the extent to which they were “biased.” 

Auditing approaches using ML algorithms have also been widely used. A situational testing auditing approach has been proposed by Luong et. al.~\cite{Luong2011}, to detect discrimination against a specific group of individuals, using an ML algorithm. K-nearest neighbors was combined with the situation testing approach to identify a group of tuples with similar characteristics to a target individual. Zhang et. al.~\cite{zhang_situation_2016} proposed an improvement over the method of Luong et. al.~\cite{Luong2011}, by engaging Causal Bayesian networks (CBNs), which are probabilistic graphical models used for reasoning and inference. For the development of a CBN, the causal structure of the dataset and the causal effect of each attribute on the decision are used to guide the identification of the similar tuples to a target individual. Robertson et. al.~\cite{robertson_auditing_2019}, present an auditing approach in the form of an opaque algorithm, called \lq{}\lq{}recursive algorithm interrogation\rq{}\rq{} used for detecting bias in search engines. The auto-complete functions of Google and Bing are treated as opaque algorithms. They recursively submitted queries, and their resulting child queries, in order to create a network of the algorithm’s suggestions. 
 
Hu et. al.~\cite{hu_auditing_2019} audited Google SERPs snippets, for evidence of partisanship where the generation of snippets is an opaque process. Moreover, Le et. al.~\cite{le_measuring_2019} audit Google News Search for evidence of reinforcing a user’s presumed partisanship. Using a sock-puppet technique, the browser first visited a political Web page, and then continued on to conduct a Google news search. The results of the audit suggested significant reinforcement of inferred partisanship via personalization. In addition, Eslami et. al.~\cite{eslami_be_2017} use a cross-platform audit technique that analyzed online ratings of hotels across three platforms, in order to understand how users perceived and managed biases in reviews.
 
In the HCI literature, auditing often involves characterizing the behavior of the algorithm from a user perspective. For instance, in Matsangidou and Otterbacher~\cite{matsangidou_what_2019}, the authors consider the inferences on physical attractiveness made by image tagging algorithms on images of people. They audited the output of four image recognition APIs on standardized portraits of people across genders and races. In a more recent work~\cite{barlas_see_2021}, the authors use auditing to understand machine behaviors in proprietary image tagging algorithms. The authors created a highly controlled dataset of people images, imposed on gender-stereotyped backgrounds. Evaluating five proprietary algorithms, they find that in three, gender inference is hindered when a background is introduced. Of the two that ``see'' both backgrounds and gender, it is the one with output that is most consistent with human stereotyping processes that is superior in recognizing gender. Another example is the work of Eslami et. al.~\cite{eslami_user_2019}, where the authors describe a qualitative study of online discussions about Yelp on the algorithm existence and opacity. The authors further enhanced the results by conducting 15 interviews with Yelp users who acted as auditors of the system, in an attempt to understand how the reviews filtering algorithm works.

Auditing approaches have also been used to detect bias in ML classification systems. For instance, in~\cite{buolamwini_gender_2018}, the authors (developers) audit three automated facial analysis algorithms to detect any gender inequalities in the classification results. They found that males were classified more accurately than females in all the three algorithms and that all the algorithms performed worst on darker female subjects.

Recently, automated methods for auditing have been introduced to detect gender or race bias in the context of online housing advertisements and search engine rankings. Asplund et al.~\cite{asplund_auditing_2020} propose the use of controlled ``Sock-puppet'' auditing techniques, which are automated systems that mimic user behavior in offline audits. They use these techniques to investigate gender-based and race-based discrimination in the context of online housing advertisements and any bias in search-result ranking. The authors use the definition of disparate impact to consider both application systems as fair or not.

\subsection{Discrimination Discovery}
%Both auditing and discrimination discovery approaches can be used in combination for detecting and understanding the discrimination bias or by choosing the most appropriate one based on the problem domain.
A common approach for discrimination discovery is to compute metrics in order to detect any direct/indirect discrimination of specific groups in the data. Examples of metrics include absolute measures, conditional measures or statistical tests~\cite{zliobaite_survey_2015}. Absolute measures define the magnitude of discrimination over a dataset by taking into account the protected characteristics and the predicted outcome. Statistical tests, rather than measuring the magnitude of discrimination, indicate the presence or absence of discrimination at a dataset level. Conditional measures compute the magnitude of discrimination that cannot be explained by any non-protected characteristics of individuals. Fairness notions have also been used in many works as metrics for discrimination. %These metrics have been used in the following works.

In Bellogin et. al.~\cite{bellogin_statistical_2017}, the authors detect statistical biases in the evaluation metrics used in recommender systems that affect the effectiveness of the recommendations. They found out that there is sparsity and popularity bias on the evaluation metrics. Many works focus on investigating the racial bias in advertising recommendations systems. For instance, Sweeney~\cite{sweeney_discrimination_2013} investigates the racial bias in advertising recommendations by an ad server when searching for particular names in Google and Reuters search engines. She finds that ads for services providing criminal records on names were significantly more likely to be served if the name search was on a typically Black first name. Ali et al.~\cite{ali_discrimination_2019}, Speicher et. al~\cite{speicher_potential_2018} and Imana et al.~\cite{imana_auditing_2021} detected significantly skewed ad delivery on racial lines in Facebook ads for employment, financial services and housing. More specifically, in~\cite{imana_auditing_2021}, the authors first build an audience that allows them to infer the gender of the ad recipients on the platforms that do not provide ad delivery statistics along gender lines, i.e., Facebook, Linkedin. They use this audience to distinguish between skew in ad delivery due to protected categories from the skew due to differences in qualifications among people in the targeted audience. Indirectly, they measure the ``equality of opportunity" fairness notion.

Another example of bias detection in RecSys and online social networks is the work of Chackraborty et al.~\cite{chakraborty_who_2017} who detect demographic bias in the input data of crowds in Twitter who make posts worthy of being recommended as trending. The bias is detected by comparing the characteristics of the trend promoters with the demographics of the general population of Twitter.
%add more details here
%Fairness measures such as disparate impact and disparate mistreatment are also used for measuring discrimination.
Apart from demographic bias, political bias is very common in social networks. In Jiang et al.~\cite{jiang_reasoning_2020}, the authors measure the fairness in social media contexts based on the fairness notions: \textit{demographic parity} and the \textit{equalized odds}. The authors detect political bias through content moderation. Bias in the social platform Facebook has also been assessed through reverse engineering of the Facebook API ranking algorithm using logistic regression in~\cite{ho_how_2020}. More specifically, the authors identify the features of a post that would affect its odds of being selected. Sentiment analysis reveals that there are significant differences in the sentiment word usage between the selected and non-selected posts.

In information retrieval systems, discrimination discovery is primarily used in user-focused studies. Weber and Castillo~\cite{weber_demographics_2010} conducted a study of user search habits, which involved a large-scale analysis of Web logs from Yahoo!. Using the logs, as well as users’ profile information and US-census information (e.g., average income within a given zip code), the authors were able to characterize the typical behaviors of various segments of the population and detect any discrimination  related to the users' sensitive demographic attributes. In a similar manner, Yom-Tov~\cite{yom-tov_demographic_2019} used search query logs to characterize the differences in the way that users of different ages, genders and income brackets, formulate health-related queries. His driving concern was the ability to discover users with similar profiles, according to their demographic information (user cohorts), who are looking for the same information e.g., a health condition. 

Pal et al.~\cite{pal_exploring_2012} considered the identification of experts in the context of a question-answering community. Their analysis revealed that as compared to other users with less expertise, experts exhibited significant selection biases in their engagement with content. They proposed to exploit this bias in a probabilistic model, to identify both current and potential experts. A method to identify selection bias, IMITATE, has also been proposed in Dost et al.~\cite{dost_your_2020}. IMITATE investigates the dataset’s probability density, then adds generated points in order to smooth out the density and have it resemble a Gaussian, the most common density occurring in real-world applications.

In a study of information exposure on the Mendeley platform for sharing academic research, Thelwall and Maflahi~\cite{thelwall_are_2015} illustrated a \textit{home-country} bias. Articles were significantly more likely to be read by users in the home country of the authors, as compared to users located in other countries. Chen et al.~\cite{chen_investigating_2018} investigated direct and indirect (implicit) gender-based discrimination in the context of resume search engines, by a system towards its users. Direct discrimination happens when the system explicitly uses the inferred gender or other attributes to rank candidates, while indirect discrimination is when the system unintentionally discriminates against users (indirectly via sensitive attributes). The results suggested that the system under review indirectly discriminates against females, however, it does not implicitly use gender as a parameter. 

Another method for detecting bias in search engine results involves the use of metrics that quantify bias in search engines~\cite{mowshowitz_measuring_2005}. A series of articles by Wikie et. al.~\cite{wilkie_retrievability_2014, wilkie_best_2014, wilkie_algorithmic_2017} and a paper of Bashir and Rauber~\cite{bashir_relationship_2011} investigates the identification retrieval bias in IR systems. Bashir and Rauber study the relationship between query characteristics and document retrievability using the TREC Chemical Retrieval track. In Wilkie and Azzopardi~\cite{wilkie_retrievability_2014}, they examined the issue of fairness vs. performance. Wilkie and Azzopardi~\cite{wilkie_best_2014} consider specific measures of retrieval bias and the correlation to the system performance. Wilkie and Azzopardi~\cite{wilkie_algorithmic_2017} consider the issue of bias resulting from the process of pooling in the creation of test sets.

A recent study~\cite{scheuerman_how_2020} detects gender and race bias in the annotation process of training data of image databases used for facial analysis. The authors found that the majority of image databases rarely contain underlying source material for how those identities are defined. Further, when they are annotated with race and gender information, database authors rarely describe the process of annotation. Instead, classifications of race and gender are portrayed as insignificant, indisputable, and apolitical.

A set of works in the HCI domain analyzes crowdsourced data from the OpenStreetMap to detect any potential biases such as gender and geographic information bias~\cite{das_gendered_2019, quattrone_theres_2015}. In a similar vein, two other studies run a crowdsourcing study to detect any bias on human versus algorithmic decision-making~\cite{green_disparate_2019, barlas_what_2019}. Green and Chen~\cite{green_disparate_2019} run a crowdsourcing study to examine the influence of algorithmic risk assessment to human decision-making, while Barlas et. al.~\cite{barlas_what_2019} compared human and algorithmic generated descriptions of people images in a crowdsourcing study in an attempt to identify what is perceived as fair when describing the depicted person. The execution of a crowdsourcing study for detecting bias has also been used in IR systems~\cite{eickhoff_cognitive_2018, lin_impact_2015}. 

Many works study the problem of bias detection in textual data using data mining methods concerning specific protected groups. The typical approach is to extract association or classification rules from the data and to evaluate these rules according to discrimination of protected groups~\cite{romei_multidisciplinary_2014, pedreschi_measuring_2009}. For instance, Datta et. al.~\cite{datta_algorithmic_2016} analyse the gender discrimination in online advertising (Google ads) using ML techniques, to identify the gender-based ad serving patterns. Specifically, they train a classifier to learn differences in the served ads and to predict the corresponding gender. Similarly, Leavy et. al.~\cite{leavy_gender_2018} detect gender bias in training textual data by identifying linguistic features that are gender-discriminative, according to gender theory and sociolinguistics. Zhao et al.~\cite{zhao_gender_2018} detect gender bias in coreference resolution systems. They introduce a new benchmark dataset, WinoBias, which focuses on gender bias. They also use a data augmentation approach that in combination with existing word-embedding debiasing techniques, removes the gender bias demonstrated in the data. Madaan et al.~\cite{madaan_analyze_2018} detect gender discrimination in movies using knowledge graphs and word embeddings after analyzing the data (using, for example, mentions of each gender in movies, emotions of the actors during the movies, occupation of each gender in the movies, screen time, etc.) In a similar vein, Ferrer et al.~\cite{ferrer_discovering_2021}propose a data-driven approach to discover and categorize language bias encoded on the vocabulary of online communities in the Reddit platform. They use word embeddings to discover the most biased words towards protected attributes, apply k-means clustering combined with a semantic analysis system to label the clusters, and use sentiment analysis to further specify biased words. Rekabsaz and his colleagues~\cite{rekabsaz_measuring_2021} also explore the detection of societal bias in word-embedding models by utilizing the first-order co-occurrence relations between the word and the representative concept words. Islam et al.~\cite{islam_debiasing_2021} introduce a collaborative filtering method to detect gender bias in social media. Their proposed method is called Neural Fair Collaborative Filtering (NFCF). They also use debiasing embeddings, and fairness interventions via penalty term.

A cluster of works in the IR domain addresses the detection of bias such as age-based bias, and text-frequency and stylistic biases in sentiment classification algorithms~\cite{diaz_addressing_2018,Rafrafi2012CopingWT,shen_darling_2018}. Other examples of detecting bias in classifiers that use sentiment analysis are the existence of offensive language or stereotyping of sensitive attributes in automated hate speech detection algorithms ~\cite{davidson_automated_2017, badjatiya_stereotypical_2019} and the detection of cultural biases at Wikipedia pages using sentiment analysis~\cite{callahan_cultural_2011}. Shandilya et al.~\cite{shandilya_fairness_2018} also detect the under-representation of sensitive attributes in the summarization algorithms. 

Keyes~\cite{keyes_misgendering_2018} identified the problem of automatic gender recognition in HCI research and how the approaches followed until recently are discriminatory towards trans gendered people. For systems to be fair, Keyes~\cite{keyes_misgendering_2018} proposed alternative methods and the development of more inclusive approaches in the gender inference process and evaluation. Apart from automatic gender recognition, an additional significant advancement in the field of HCI is that of data-driven personas. Salminen et al.~\cite{salminen_detecting_2019} investigated the presence of demographic bias in automatically generated, data-driven personas. They discovered that the more personas they generated, the more diverse the sample became in terms of gender and age representation. Practitioners who use data generated personas should consider the possibility of unintentional bias in the data they use, that consequently is transferred to the personas they generate.  

Multiple other approaches have been proposed in ML literature that detect any discrimination in the data or classifier. Choi et al.~\cite{choi_learning_2020} discover and mine discrimination patterns that refer to whether an individual is classified differently depending on whether some sensitive attributes were observed. The algorithm detects discrimination patterns in a Naive Bayes classifier using branch and bound search and removes them. It learns maximum likelihood parameters based on these parameters. Pedreshi et al.~\cite{pedreschi_measuring_2009} use an opaque predictive model to extract frequent classification rules based on an inductive approach. Background knowledge is used to identify the groups to be detected as potentially discriminated. On the other hand, Zhang et. al.~\cite{zhang_causal_2017} use a causal Bayesian network and a learning structure algorithm to identify the causal factors for discrimination. The direct causal effect of the protected variable on the dependent variable represents the sensitivity of the dependent variable to changes in the discrimination grounds while all other variables are held fixed. They also detect discrimination in the prediction/classification outcome by computing the classification error rate (error bias). In a more recent work, Zucker et al.~\cite{zucker:2020} introduce a new domain-specific programming language, the Arbiter for ML practitioners. It allows users to make guarantees about the level of bias in any produced models. 

The notion of divergence \cite{pastor:2021}, which estimates the difference in classification performance measures, has also been proposed as a metric to identify data subgroups in which a classifier performs differently. Pastor et al.~\cite{pastor_how_2021} introduce the DivExplorer, an interactive visual analytics tool that identifies algorithmic bias using the divergence notion. An interactive system to detect fairness issues in the classifiers has also been proposed in~\cite{li_denouncer_2021}. The system is called DENOUNCER and it allows users to explore fairness issues for a given test dataset, considering different fairness notions. In addition, Nargesian et al.~\cite{nargesian_tailoring_2021} detect the groups in the dataset that are unfairly treated by the classifier by developing an exploration-exploitation based strategy. Their approach captures the cost and approximations of group distributions in the given dataset.
%\subsubsection{Direct Discrimination Detection of Perception Bias}

In IR systems, a common type of bias is the cognitive or perception bias that arises from the manner in which information is presented to users, in combination with the user's own cognition and/or perception. For example, Jansen and Resnick~\cite{jansen_examination_2006} analyzed the behaviors of 56 participants engaged in e-commerce search tasks, with the goal of understanding users’ perceptions of sponsored versus un-sponsored (organic) Web links. The links suggested by the search engine were manipulated in order to control content and quality. Even controlling for these factors, it was shown that users have a strong preference for organic Web links. In a similar vein, Bar-Ilan et al.~\cite{barilan_presentation_2009} conducted a user experiment to examine the effect of position in a search engine results page (SERPs). Across a variety of queries and synthetic orderings of the results, they demonstrated a strong placement bias; a result’s placement, along with a small effect on its source, is the main determinant of perceived quality. User perception is also examined in a study~\cite{metaxa_image_2021} where the authors consider people's impressions of occupations and sense of belonging in a given field when shown search results with different proportions of women and people of color. They find that both axes of representation as well as people's own racial and gender identities impact their experience of image search results. Gezici et al.~\cite{gezici_evaluation_2021} propose a new evaluation framework to measure bias in the content of SERPs (on political and controversial search topics) by measuring stance and ideological bias. They propose three novel fairness-aware measures of bias based on common IR utility-based evaluation measures.

Ryen White, of Microsoft Research, has published extensively on detecting users' perception bias during and after a search, particularly when trying to find information to answer health-related queries. In an initial work~\cite{white_belief_2014}, a user study focused on finding yes-no answers to medical questions, showed that pre-search beliefs influence users' search behaviors. For instance, those with strong beliefs pre-search, are less likely to explore the results page, thus reinforcing the above-mentioned positioning bias. A follow-up study by White and Horvitz~\cite{white_belief_2015} looked more specifically at users’ beliefs on the efficiency of medical treatments, and how these beliefs could be influenced by a Web search. An example of searching for user perception bias in recommender systems was presented in~\cite{rosenblat_algorithmic_2016}, where drivers' perceptions of the Uber application were investigated, taking into consideration drivers' profiles and their history performance.

\subsection{Bias Detection Comparison}

%The first step in mitigating bias in an algorithmic system \green{is to detect the source of bias, the protected group and their protected characteristics based on which they are discriminated (i.e. sensitive/protected attributes).}  
Table~\ref{tab:comp:discri} summarizes the methods used for auditing and discrimination discovery within each of the research domains analyzed in this survey. In ML systems, bias detection is mostly done using discrimination or fairness metrics. Auditing in ML systems can be achieved by auditing software tools or when \textit{developers/regulators} act as auditors of the algorithmic system. On the other hand, in IR, HCI and RecSys systems, \textit{users} often act as auditors by submitting different queries in search engines and social networks or by taking the role of crowdworker in the crowdsourcing conducted studies. Discrimination discovery approaches used in IR, HCI and RecSys systems are similar to auditing but with a more concrete methodology on detecting bias.

\begin{table}[h]
\small
\begin{tabular}{|l|l|l|l|}
\hline
 \textbf{Domain} &  \textbf{Problem} & \textbf{Solution Space} &   \textbf{Reference(s)} \\
 \hline
 \multicolumn{4}{|c|}{\textbf{Bias Detection}}\\\hline
 ML & Data/Model & Auditing & Automatic auditing tool \cite{saleiro_aequitas:_2018, bellamy_ai_2018} \\
 &&&Developers as auditors \cite{buolamwini_gender_2018, Luong2011, zhang_causal_2017}\\
 & Data & & Discrimination/Fairness metrics~\cite{zliobaite_survey_2015, zafar:2017, jiang_reasoning_2020} \\
 &Data && Metrics\cite{dost_your_2020, pastor_how_2021, li_denouncer_2021}\\
 & Data/Model/Output & Discrimination Discovery & ML methods  \\&&&\cite{leavy_gender_2018, zhang_causal_2017, pedreschi_measuring_2009, datta_algorithmic_2016, cowgill_algorithmic_2017, zucker:2020}\\
 \hline
 IR & User/Data/Output & Auditing & Submit queries to search engines/ Twitter\\ &&& \cite{vincent_measuring_2019, kay_unequal_2015, magno_stereotypes_2016, otterbacher_competent_2017,kulshrestha_quantifying_2017, hu_auditing_2019, le_measuring_2019}\\ 
 &Model/User& & Sock-puppet auditing\cite{asplund_auditing_2020}\\
 & User/Data/Output & Discrimination Discovery & Analysis of Web logs\\&&& \cite{weber_demographics_2010, yom-tov_demographic_2019, pal_exploring_2012, chen_investigating_2018, wilkie_retrievability_2014,wilkie_best_2014, wilkie_algorithmic_2017,bashir_relationship_2011, imana_auditing_2021}\\
 &User/Data/Output&&Word embedding\cite{ferrer_discovering_2021, rekabsaz_measuring_2021, islam_debiasing_2021}\\
&User/Third Party/Data && Crowdsourcing studies\cite{eickhoff_cognitive_2018, lin_impact_2015}\\
&User/Third Party && Direct discrimination of perceived bias \\&&&\cite{jansen_examination_2006, barilan_presentation_2009, white_belief_2014, white_belief_2015}\\
          \hline
HCI & Output/Model/User & Auditing& Analysing system behavior\\&&&\cite{matsangidou_what_2019, johnson_effect_2017}\\
& Data/User/Third Party & Discrimination Discovery& Crowdsourcing studies \\&&&\cite{das_gendered_2019,quattrone_theres_2015, green_disparate_2019, barlas_what_2019, metaxa_image_2021}\\
&Model/User && Use of ML methods\cite{ho_how_2020, scheuerman_how_2020}\\
&Data/User&&Data-driven personas\cite{salminen_detecting_2019}\\
\hline
RecSys & Data/User & Auditing & Developers as auditors\\&&& \cite{eslami_be_2017, edelman_racial_2017}\\
 &Model/User & &Sock-puppet auditing\cite{asplund_auditing_2020}\\
& User/Model/Output & Discrimination Discovery & Discrimination detection in advertising\\&&& recommendation systems\cite{ali_discrimination_2019, speicher_potential_2018, sweeney_discrimination_2013}\\
        & Output/Model & & Discrimination detection in evaluation metrics\\&&& \cite{bellogin_statistical_2017, ekstrand_all_2018}\\
        &Output/User&&Discrimination in social networks \cite{chakraborty_equality_2019}\\
        \hline        
\end{tabular}
\caption{Comparison of Discrimination Detection approaches across the four domains.}
\label{tab:comp:discri}
\end{table}
%\begin{figure}
%    \centering
%    \includegraphics[width=0.8\textwidth]{diversity_dimensions_accross_domains.png}
%    \caption{Diversity dimensions where discrimination has been detected in the surveyed papers across different domains}
%    \label{fig:diversity:discrimination}
%\end{figure}

%Fig.~\ref{fig:diversity:discrimination} analyzes the frequency with which specific protected characteristics were examined in the literature across domains. The information dimension has clearly been the most studied dimension in the FAT literature thus far. As mentioned, information is the primary dimension addressed in the ML literature (in particular, with respect to explainability). Likewise, IR articles often consider information as the diversity dimension under study; here, the classic example is the large body of work on search engine biases. In contrast, the literature in HCI and RecSys do not often address information as a diversity dimension. In these fields, FAT-related articles more often consider social and cultural dimensions.

\section{Fairness Management}\label{sec:fairness}
The second set of tools used in mitigating algorithmic system bias concerns processes of \textit{Fairness Management}. One consideration is to use fairness management approaches to mitigate the bias detected in any part of an algorithmic system. However, in order to make sure that an algorithmic system can be considered ``fair,'' it is not enough to simply mitigate the detected bias -- the design of the system should be ``fairness-aware.'' In this section, we provide details of the solution approaches proposed in the literature in each of the five fairness management categories.

\subsection {Fairness Pre-processing Methods}
%Many of the articles that concern the discrimination discovery and fairness problems use pre-processing methods to remove the discriminatory bias of the input training data. 
An approach that is usually used to mitigate bias on the input or training data is the removal of sensitive attributes that may be involved in discrimination. However, in some cases, the inclusion of sensitive characteristics in the data may be beneficial to the design of a fair model~\cite{zliobaite_using_2016}. To handle this issue, some approaches remove information about the protected variables from the training data but they also transform the training data using data mining methods. For instance, Kamiran and Calders~\cite{kamiran2009classifying} use a naive Bayes classifier to generate rankings of each observation in the training data based on its probability of belonging to the desirable class category. The outcome variable in the training data is adjusted until there is no remaining association between the protected variable and the intended outcome variable. The drawback of this solution is that it is limited to a binary outcome variable and the transformed training data cannot be used with other outcome variables. 

Calders and Verwer~\cite{calders2010three} eliminate the above drawbacks by presenting three algorithms that transform (i.e., re-weight) the training data based on an objective function that is minimized when the outcomes from a model that fit to the transformed data are independent of the protected variable. This class is also restricted to binary outcome and protected variables. Data transformation approaches have also been proposed by Johndrow and Lum~\cite{johndrow_algorithm_2019} and Zemel~\cite{zemel_learning_2013}. Johndrow et al.~\cite{johndrow_algorithm_2019} suggest a statistical framework where the training data are transformed by mapping individuals to an input space that is mutually independent of specific groupings. In~\cite{zemel_learning_2013}, the authors encode the data by mapping each individual, represented as a data point in a given input space, to a probability distribution in a new representation space. The aim of this is to hide any sensitive information that can identify if the individual belongs to a protected group. Percy et al.~\cite{percy2020lessons} propose an approach to mitigate gender bias on gambling. The method uses gender data for training only, constructing separate models for each gender and combining trained models into an ensemble that does not require gender data once deployed.

Another frequent pre-processing technique is the use of directed acyclic graphs and causal reasoning that capture the dependencies between the features and their impact on the outcome. For instance, Zhang et al.~\cite{zhang_causal_2017} discover and prevent discrimination bias in decision support systems using a causal Bayesian network (BN) to identify pairs of tuples with similar characteristics from the dataset. Then, they generate a new dataset sampled from the learned BN. Cardoso et al.~\cite{l._cardoso_framework_2019} also use a Bayesian network estimated from real-world data to generate biased data that are learned from real-world data. A data transformation method has also been applied to ensure fairness in RecSys~\cite{wu_learning_2021}. The authors propose a new graph-model technique, the FairGo model, which ensures fairness for every recommender system by transforming the original embedding of user and items into a filtered embedding space based on the sensitive feature set. FairGo is model-agnostic and can be applied to multiple sensitive attributes.

Rather than adjusting/transforming the observations of the training data, other works use re-labeling. Cardoso et al.~\cite{l._cardoso_framework_2019} propose the use of an auditing tool to repair the dataset by changing attribute labels. Kamiran and Calders~\cite{kamiran2009classifying} massage the data by swapping some of the labels in such a way that a positive outcome for the disadvantaged group is more likely and then they re-train the model. Feldman et al.~\cite{feldman_certifying_2015} proposed the \textit{disparate impact removal} solution approach that manipulates individual data dimensions in a way that depends on the protected attribute.
 
 Similar techniques to data transformation, but that consider the selection of features, have been introduced in~\cite{salazar_automated_2021, celis_implicit_2020}. Salazar et al.~\cite{salazar_automated_2021} use a multi-objective optimization algorithm for feature construction. They use this approach to generate more features that lead to both high accuracy and fairness by applying human understandable transformations. Celis et al.~\cite{celis_implicit_2020} develop a novel approach that takes as input a visibly diverse control set of images of people and uses this set as part of a procedure to select a set of images of people in response to a query. The goal is to have a resulting set that is more visibly diverse in a manner that emulates the diversity depicted in the control set.

Other popular fairness pre-processing methods are the re-sampling methods that generate a balanced dataset that will not under- or over-represent a particular protected group~\cite{dixon_measuring_2018,shen_darling_2018, diaz_addressing_2018, johnson_effect_2017, gjoka_walking_2010}. A similar approach is used in RecSys where a re-sampling method is used to balance the neighborhoods before producing recommendations~\cite{burke2018balanced} or re-balance the input data according to the protected attributes (e.g., gender) to produce a fair training dataset~\cite{Luong2011}. A re-sampling method has also been used by Sharma et al.~\cite{sharma_data_2020}. They use a data augmentation technique that adds synthetic data for removing bias in the data. The technique selectively adds only a subset of the synthetic points to create new augmented dataset to meet the fairness criteria while maintaining accuracy. %Another approach used in RecSys is to make recommended items independent from protected info~\cite{Kamishima 2017}.

%Ekstrand et. al.~\cite{ekstrand_all_2018} researched the effect of demographics on evaluation metrics in recommender systems.

%The aim of fairness sampling methods is to generate a balanced dataset that will not under- or over-represent a particular protected group. Fairness sampling is usually achieved by re-balancing the data using various data re-sampling techniques that can be applied to any type of algorithmic system~\cite{dixon_measuring_2018,shen_darling_2018, diaz_addressing_2018, johnson_effect_2017, gjoka_walking_2010}. In recommender systems, an example of fair sampling approach is to balance the neighborhoods before producing recommendations or re-balance the input data according to the protected attributes e.g., gender~\cite{Luong2011}. 

 %Cardoso et al. (2019) propose the use of a black-box auditing tool to repair the dataset by changing attribute labels. %Similarly, Pedreshi et al. (2009) use a black-box predictive model to extract frequent classification rules based on an inductive approach. Background knowledge is used to identify the groups to be detected as potentially discriminated. In addition, 

\subsection{Fairness In-processing Methods}
%\textit{[to add an introductory paragraph...]} %Therefore, the methods modify the classification/predictive algorithm mainly by introducing some fairness constraints~\cite{zhang_fairness_2018, celis_classification_2019, kleinberg_inherent_2016, dimitrakakis_bayesian_2018} or by introducing new fairness metrics such as FACE and FACT~\cite{khademi_fairness_2019}, feature-apriori fairness, feature accuracy fairness and feature-disparity fairness~\cite{grgic-hlaca_human_2018}. 
%Wu et al.~\cite{wu_convexity_2019} propose a framework that uses many of these fairness metrics as convex constraints that are directly incorporated into the classifier. They first present a constraint-free criterion (derived from the training data) which guarantees that any learned classifier will be fair according to the specified fairness metric. Thus, when the criterion is satisfied, there is no need to add any fairness constraint into optimization for learning fair classifiers. When the criterion is not satisfied, a constrained optimization problem is used to learn fair classifiers. 

One category of the in-processing approaches is the use of an optimization method. Xiao et al.~\cite{xiao_fairness-aware_2017} suggest a multi-objective optimization framework optimizing fairness and social welfare simultaneously on group recommendation. The goal was to maximize the satisfaction of each group member while minimizing the unfairness between them. The results show that considering fairness in group recommendation can enhance the recommendation accuracy. A multi-objective optimization approach has also been proposed in~\cite{nguyen_approximate_2020} for fair allocations using two criteria, maximum fairness and efficiency. They propose a dynamic programming algorithm to construct an appropriate Pareto set. %Hu et al.~\cite{hu_auditing_2019} use fairness constraints to formulate the fair classification problem as a constrained optimization problem. They learn multiple fair classifiers simultaneously from a single training dataset.} 

Optimization approaches with fairness weights have also been used in recommender systems for two-sided marketplaces~\cite{mehrotra_towards_2018}. In that scenario, the developed recommendation systems should be fair on both the demand and supplier sides. The authors propose different recommendation policies that jointly optimize the relevance of recommendations to consumer (i.e., user) and fairness of representation of suppliers. Kusner et al.~\cite{kusner_counterfactual_2017} focus on satisfying the counterfactual fairness as the notion of fairness. They capture the social biases that may arise towards individuals based on sensitive attributes. They provide optimization of fairness and prediction accuracy of the classifier using a causal model.
%[Move to post-processing] Another suggestion for a fair ranking system proposed by~\cite{chakraborty_who_2017} who attempt to find a fair ranking for crowd-sourced recommendations taking into account that the vast majority of potential voters are silent, that some people vote multiple times, and that votes for similar topics are split, leading to a bias towards extreme viewpoints. Regarding the fairness of image ranking recommendations, Karako and Manggala~\cite{karako_using_2018}  incorporated fairness in the system by choosing a sample of labeled images, based on gender, though their method is suitable for any attribute of interest. 

A second category of in-processing methods is the use of regularization methods. Yan and Howe~\cite{yan_fairness-aware_2020} introduce the FairST, a fairness-aware demand prediction model for spatiotemporal urban applications. 
Two spatio-temporal fairness metrics have been introduced as a form of regularization to reduce discrimination in demographic groups. Kamishima et al.~\cite{kamishima_fairness-aware_2012}, also use a regularization approach that can be applied to any algorithmic model (classifier). They introduce a prejudice remover regularizer that enforces classifier’s independence from sensitive attributes. 

Instead of applying a regularization method, Rezaeil et al.~\cite{rezaei_fairness_2020} mitigate bias detected in any classifier by re-building the classifier and incorporating fairness constraints to the predictor. The method reshapes the predictions (output) for each group to satisfy the fairness constraints that consider the protected groups.

In ranking systems, IR and RecSys, in-processing approaches primarily explore the mitigation of bias in the ranking algorithms using learn-to-rank methods. For instance, Dai and his colleagues~\cite{dai_adversarial_2021} propose a novel framework, Adversarial Imitation Click Model (AICM), which is based on imitation learning to address the exposure bias in click-models. Click-models rely on learning-to-rank, by studying how users interact with a ranked list of items. In~\cite{ovaisi_correcting_2020}, the authors consider both the selection and position bias in the rank-based results. They frame the problem as a counterfactual problem and adapt Heckerman's (rank) approach by combining it with position bias correction methods to correct both the selection and position bias. Yang and Ai~\cite{yang_maximizing_2021} propose a fair and unbiased ranking method named Maximal Marginal Fairness (MMF) for dynamic learning to rank, to achieve both fairness and relevance in top-k results.

In a recent work~\cite{kuhlman_rank_2020}, the authors introduce a fair
rank aggregation framework for aggregating multiple rankings in a database, which can be applied to the databases of the ranking systems. It uses pairwise discordance to both compute closeness among consensus and base rankings and measure the advantage given to each group of candidates. Another fairness issue in ranking systems concerns the mitigation of bias in the PageRank algorithm. The authors in~\cite{tsioutsiouliklis_fairness-aware_2021} provide a parity-based definition of fairness that imposes constraints on the proportion of PageRank allocated to the members of each group. They validate the fairness notion of local and personalized PageRank fairness. 
%An in-processing method that has been proposed in information retrieval systems considers the interaction between the user and a system, or a particular system component, as possible insight in solving information biases. Mitra et al.~\cite{mitra_user_2014} presented the first large-scale study of users’ interaction with the auto-complete function of Bing. Through an analysis of query logs, they found evidence of a position bias (i.e., users were more likely to engage with higher-ranked suggestions). They were also more likely to engage with auto-complete suggestions after having typed at least half of their query. In a follow-up study, Hofmann et al.~\cite{hofmann_eye-tracking_2014} conducted an eye-tracking study with Bing users. In half of their queries, users were shown ranked the auto-complete suggestions whilst in the other half of queries, the suggestions were random. The authors confirmed the position bias in the auto-complete results, across both ranking conditions. They found that the quality of the auto-complete suggestions affected search behaviours; in the random setting users visited more pages in order to complete their search task.

%In the HCI domain, researchers use a human-in-the-loop approach for decision-making when sensitive attributes are involved rather than the statistical model approach~\cite{brown_toward_2019}. 

\subsection{Fairness Post-processing Methods}
The most well-known post-processing method used in ML literature is the re-labeling of the decision outcome. Pedreschi et al.~\cite{pedreschi_measuring_2009} alter the confidence of classification rules inferred by the CPAR algorithm, whereas Kamiran et al.~\cite{kamiran2009classifying} re-label the class that is predicted at the labels of a decision tree. In~\cite{hardt_equality_2016}, the authors propose a new fairness definition to optimally adjust any learned predictor so as to remove discrimination. Their framework constructs classifiers from any Bayes optimal regressor following a post-processing step that minimizes the loss in utility. 

Additionally, in the literature of rank-based systems (i.e. IR and RecSys), post-processing methods focus on the re-ranking of the recommended or search results. In~\cite{li_user-oriented_2021}, the authors provide a re-ranking approach to mitigate the unfairness problem between active and inactive users by adding constraints over evaluation metrics. Experiments show that their approach improves group fairness of users in recommender systems, and also achieves better overall recommendation performance. 
  
%[Move to pre-processing or discirmination discovery] A method proposed by Chackraborty et al. ~\cite{chakraborty_who_2017} attempt to find a fair ranking for crowd-sourced recommendations taking into account that the vast majority of potential voters are silent, that some people vote multiple times, and that votes for similar topics are split, leading to a bias towards extreme viewpoints.
A re-ranking method has also been proposed by Karako and Manggala~\cite{karako_using_2018} who introduce a fairness-aware variation of the Maximal Marginal Relevance (MMR). The proposed method incorporates fairness in a recommender or search system by choosing a sample of labeled images, based on gender when retrieving untagged images similar to an input image or query. Mitigation of gender bias on image tagging has been proposed in~\cite{tang_mitigating_2021} where the authors introduce the Guided Attention Image Captioning model (GAIC). The GAIC pipeline encourages the model to provide correct gender identification with high confidence when gender evidence in image is obvious. When gender evidence is vague or occluded, GAIC tends to describe the person with gender neutral words, such as ``person'' and ``people.'' In addition, Zehlike et al.~\cite{zehlike_fair:_2017} and Singh and Joachims~\cite{singh_fairness_2018} propose fair top-k ranking algorithms for RecSys that makes the recommendations subject to group fairness criteria and constraints. 

``Other'' works in ranking systems propose methods to achieve the fairness of the general ranking results, rather than focusing on the top-k ranking. Patro and colleagues~\cite{patro_fairrec_2020} propose the FairRec algorithm, which validates exhibiting the desired two-sided fairness, both consumer and producer fairness, by mapping the fair recommendation problem to a fair allocation problem. Kuhlman et al.~\cite{kuhlman_fare:_2019} use an auditing methodology FARE (Fair Auditing based on Rank Error) for error-based fairness assessment of the ranking results. They propose three error-based fairness criteria, which are rank-appropriate, to assess the correctness of the rankings. In addition, Kirnap et al.~\cite{kirnap_estimation_2021} estimate four fair ranking metrics by acquiring group membership annotations for a sample of documents in its corpus.

%The fairness notions taken into consideration in this study are the ones defined in RecSys: envy-freeness (EF) and proportional-fair-share (PFS).
  
%[Move this to fairness certification] 
\subsection{Fairness Certification}
Fairness certification methods are used to certify the fairness of the system using some constraints~\cite{zhang_fairness_2018, celis_classification_2019, kleinberg_inherent_2016, dimitrakakis_bayesian_2018} or by introducing new fairness metrics such as FACE and FACT~\cite{khademi_fairness_2019}, feature-apriori fairness, feature accuracy fairness and feature-disparity fairness~\cite{grgic-hlaca_human_2018}. Wu et al.~\cite{wu_convexity_2019} propose a framework that uses many of these fairness metrics as convex constraints that are directly incorporated into the classifier. They first present a constraint-free criterion (derived from the training data) that guarantees that any learned classifier will be fair according to the specified fairness metric. Thus, when the criterion is satisfied, there is no need to add any fairness constraint into the classifiers. When the criterion is not satisfied, a constrained optimization problem is used to learn fair classifiers.

Hu et al.~\cite{hu:2020} propose a metric-free individual fairness based on the gradient contextual bandit algorithm that aims to maximize fairness. In~\cite{galhotra_reliable_2020}, the authors use multi-objective clustering algorithms to maximize both accuracy and fairness and also to introduce diversity and transparency as constraints.
Counterfactual fairness is another well-known metric used for certifying the fairness of the system. In~\cite{sharma_certifai_2020}, counterfactual explanations evaluate fairness with respect to a particular individual as well as the fairness of the model towards groups of individuals. They define the metric ``burden'' to evaluate group fairness. The burden is computed taking into consideration how much the fitness measure differs for counterfactuals generated for specific groups of individuals. Cruz Cortes et al.~\cite{cruz_cortes_invitation_2020} use population fairness metrics: Predictive Parity and Error Rate Balance. They propose a simple agent-based model to detect any discrimination inequalities in an arrest-sentence system. Group fairness has been used as a definition in recommender systems for group recommendations as well. Kaya et al.~\cite{kaya:2020} define a new metric for group fairness called Group Fairness Aware Recommendations (GFAR) considering the fairness of the top-N ranked items. GFAR defines top-N ranking as fair when the relevance of each of the top-N items, to the group members is ‘balanced’ across the group members. 
 
In information retrieval systems, researchers often focus on user evaluation to certify the fairness of the system. Mitra et al.~\cite{mitra_user_2014} presented the first large-scale study of users’ interactions with the auto-complete function of Bing. Through an analysis of query logs, they found evidence of a position bias (i.e., users were more likely to engage with higher-ranked suggestions). They were also more likely to engage with auto-complete suggestions after having typed at least half of their query. In a follow-up study, Hofmann et al.~\cite{hofmann_eye-tracking_2014} conducted an eye-tracking study with Bing users. In half of their queries, users were shown the ranked auto-complete suggestions while in the other half of queries, the suggestions were random. The authors confirmed the position bias in the auto-complete results, across both ranking conditions. They found that the quality of the auto-complete suggestions affected search behaviors; in the random setting users visited more pages in order to complete their search task. Another popular fairness certification method is simply to raise users' awareness. Epstein et al.~\cite{epstein_suppressing_2017} develop solutions for the Search Engine Manipulation Effect (SEME), citing recent evidence of its impact on the views of undecided voters in the political context. In a large-scale online experiment with 3,600 users in 39 countries, they showed that manipulating the rankings in political searches can shift users’ expressed voting preferences by up to 39\%. However, providing users with a “bias alert,” which informed them that “the current page of search rankings you are viewing appears to be biased in favor of [name of candidate],” reduced the shift to 22\%. They found that this could be reduced even further when more detailed bias alerts were provided to users. Nonetheless, they reported that SEME cannot be completely eliminated with this type of intervention, and suggested instituting an “equal-time” rule such as that used in traditional media advertisements.

``Other'' works use alternative approaches rather than the computation of specific metrics to certify an ML system. For instance, Fang et al.~\cite{fang_achieving_2020} certify the fairness of a classifier by constructing fairgroups, considering the feature importance to the decision variable. Individuals with similar features are grouped into clusters. This approach adopts the notion of fairness related to disparate impact, which affects individuals with at least one protected feature. In addition, Kilbertus et al.~\cite{kilbertus_blind_2018} provide fairness learning and certification without access to users' sensitive data. To achieve this, they use an encrypted version of sensitive data, privacy constraints and decision verification by employing secure multi-party computation (MPC) methods. The use of techno-moral graphs for certifying ML algorithmic systems was also suggested in~\cite{jaton_assessing_2021}. The authors argue that a three-dimensional conceptual space can be used to map machine learning algorithmic projects in terms of the morality of their respective and constitutive ground-truth practices. Such techno-moral graphs may, in turn, serve as equipment for greater governance of machine learning algorithms and systems.

\subsection{Fairness Perception}
Woodruff et al.~\cite{woodruff_qualitative_2018} explore, in a qualitative study, the perception of algorithmic fairness by populations that have been marginalized. In particular, they consider how race and low socioeconomic status was used in stereotyping and adapting services to those involved. Most participants were not aware of algorithmic unfairness even though they had experienced discrimination in their daily lives. Brown et al.~\cite{brown_toward_2019} also present a qualitative study for understanding the public’s perspective on algorithmic decision-making in public services. They discovered that many participants mentioned discrimination and bias based on race, ethnicity, gender, location, and socioeconomic status. A descriptive approach for identifying the notion of perceived fairness for machine learning was suggested by Srivastava et al.~\cite{srivastana:2019}. They argued that the perceived fairness of the user is the most appropriate notion of algorithmic fairness. Their results show that the formal measurement, demographic parity, most closely matches the perceived fairness of the users and that in cases when the stakes are high, accuracy is more important than equality. 

Perceived fairness on algorithmic decision-making is also explored in~\cite{wang_factors_2020} where the authors conduct an online experiment to better understand perceptions of fairness, focusing on three sets of factors: algorithm outcomes, algorithm development and deployment procedures, and individual differences. They find that people rate the algorithm as more fair when the algorithm predicts in their favor, even surpassing the negative effects of describing algorithms that are very biased against particular demographic groups. This effect is moderated by several variables, including participants’ education level, gender, and several aspects of the development procedure. These findings suggest that systems that evaluate algorithmic fairness through users’ feedback must consider the possibility of ``outcome favorability'' bias.

In another study, the authors identify perception bias in borderline fact-checking messages~\cite{park_experimental_2021}. The authors conduct both a quantitative and qualitative study by conducting semi-supervised user interviews to learn the user experience and perception of different fact-checking conditions. In a recent work~\cite{masrour_fairness_2020}, the authors introduce a network-centric fairness perception function that can be viewed as a local measure of individual fairness.

In addition, Maxwell et al.~\cite{maxwell_impact_2019} investigated the influence of result diversification on users’ search behaviors. Diversification can reduce search engine biases by exposing users to a broader coverage of information on their topic of interest. A within-subject study with 51 users was performed, using the TREC AQUAINT collection. Two types of search tasks - ad hoc versus aspectual - are assigned to each user using a non-diversified IR system as well as a diversified system. Results indicated significant differences in users’ search behaviors between the two systems, with users executing more queries, but examining fewer documents when using the diversified system on the aspectual (i.e., more complex) task. 

\subsection{Fairness Management Comparison}
%Fairness management approaches can be classified into pre-processing, in-processing and post-processing methods. Pre-processing methods handle bias in input data, in-processing methods concern the mitigation of bias in the algorithm and post-processing methods concern the elimination of bias in the outcome. 
As displayed in Table~\ref{tb:fairness:comparison}, in ML algorithmic systems, the most popular techniques are data re-sampling, removal of sensitive attributes and data transformation to mitigate bias in the data, optimization and regularization approaches to mitigate bias during the model training and re-labeling of the outcome decision to mitigate bias on the output of the system. In ranking systems such as RecSys and IR, the most popular approaches are re-sampling for mitigating data bias, learning to rank methods to mitigate bias in the ranking algorithms and re-ranking methods as for modifying the ranking outcomes. Two approaches that are common in RecSys and ML communities are the data transformation (fairness pre-processing) and optimization approaches (fairness in-processing). In the HCI community, since the \textit{user} is the main stakeholder, most of the papers examine the user perception on fairness. Approaches to mitigate bias referred to the use of a human-in-the-loop on the decision-making~\cite{brown_toward_2019}.  
Fairness certification techniques use fairness constraints or defining new fairness notions, i.e., counterfactual fairness and metrics for certifying the fairness of systems in all the four research domains. In IR, some studies also use user evaluation to certify the fairness of the system. 
%\green{Even if there are only few works studying the fairness perception across the domains, % In IR, RecSys and HCI systems use mostly pre-processing approaches such as re-sampling methods, removal of sensitive attributes and feature selection to handle bias in data and \green{fairness certification approaches i.e. raising users' awareness of the algorithm's behavior. In the communities of RecSys and IR, researchers also proposed some post-processing methods to handle bias in the ranking results.}

\begin{table}[ht]
\small
\begin{tabular}{|l|l|l|l|}
\hline
\multicolumn{4}{|c|}{\textbf{Fairness Management}}\\\hline
  \textbf{Domain} & \textbf{Problem} & \textbf{Solution Space} & \textbf{Reference(s)} \\
         \hline
        ML & Data & Fairness Pre-processing  & Removal of protected attributes 
        \\ && & \& Data Transformation \cite{calders2010three, johndrow_algorithm_2019, zemel_learning_2013, percy2020lessons}\\
         &  & & Causal BN\cite{zhang_causal_2017, l._cardoso_framework_2019}\\
         &&&Data Re-labeling \cite{kamiran2009classifying, feldman_certifying_2015}
        \\
        &&& Re-sampling methods \cite{johnson_effect_2017, sharma_certifai_2020}

     %   ML & Model/Third Party & Fairness In-processing  & Fairness Constraints \cite{zhang_fairness_2018, celis_classification_2019, kleinberg_inherent_2016, dimitrakakis_bayesian_2018}
     \\
        %& Model/Output & &  Fairness Metrics \cite{khademi_fairness_2019, grgic-hlaca_human_2018, wu_convexity_2019}\\
        & Model&Fairness In-processing& Regularization approach \cite{kamishima_fairness-aware_2012, yan_fairness-aware_2020}\\
        &&&Optimization approach \cite{nguyen_approximate_2020, salazar_automated_2021}\\
        &&&Constraints\cite{rezaei_fairness_2020}\\
        &Model/Output&& Counterfactual fairness \cite{kusner_counterfactual_2017}\\
         & Third Party/Output & Fairness Post-processing & Altering of labels \cite{kamiran2009classifying, hardt_equality_2016, pedreschi_measuring_2009} \\
       & User/Third Party& Fairness Perception& \cite{srivastana:2019, masrour_fairness_2020}\\
       &Data/Model/Output& Fairness Certification & Fairgroups \cite{fang_achieving_2020}\\
       &&&Counterfactual Fairness \cite{kilbertus_blind_2018, sharma_certifai_2020}\\
       &&&Techno-moral graphs\cite{jaton_assessing_2021}\\
       &&&Fairness Constraints/Metrics\\&&&\cite{zhang_fairness_2018, celis_classification_2019, kleinberg_inherent_2016, dimitrakakis_bayesian_2018, khademi_fairness_2019, grgic-hlaca_human_2018, wu_convexity_2019, cruz_cortes_invitation_2020}\\
       \hline
     %   &Data/Model&& Encrypted version of sensitive data \cite{kilbertus_blind_2018}\\
       
        IR &Data &Fairness Pre-processing& Data sampling \cite{gjoka_walking_2010, dixon_measuring_2018, shen_darling_2018, diaz_addressing_2018}\\
        & Model & Fairness In-processing & Learn-to-rank methods\\&&&\cite{dai_adversarial_2021, ovaisi_correcting_2020, yang_maximizing_2021, kuhlman_rank_2020}\\
        & Output & Fairness Post-processing & Re-ranking\cite{karako_using_2018, kirnap_estimation_2021, kuhlman_fare:_2019, li_user-oriented_2021}\\
        & Model/Output/User &Fairness Certification &\cite{epstein_suppressing_2017, hofmann_eye-tracking_2014, mitra_user_2014}\\
        &User/Output& Fairness Perception&\cite{park_experimental_2021, maxwell_impact_2019}\\
        \hline 
        HCI & Data& Fairness Pre-processing & Data sampling \cite{johnson_effect_2017} \\
        &&&Data transformation \cite{celis_implicit_2020}\\
        & Output & Fairness Perception& Human-in-the-loop \cite{brown_toward_2019}\\
        &User/Output &&Metrics\cite{wang_demonstration_2021}\\
        &Output/User&Fairness Certification &\cite{woodruff_qualitative_2018, lee_algorithmic_2017}\\
        \hline
        RecSys & Data &Fairness Pre-processing & Data sampling \cite{burke2018balanced, Luong2011, karako_using_2018}\\
        & & & Data transformation \cite{wu_learning_2021}\\
       % &Model && Fair top-k ranking algorithm \cite{zehlike_fair:_2017, singh_fairness_2018, chakraborty_who_2017}\\
        &Model/Output && Optimization approaches \cite{xiao_fairness-aware_2017, mehrotra_towards_2018} \\
        &Model & Fairness In-processing & Learn-to-rank~\cite{yang_maximizing_2021, kuhlman_rank_2020} \\
       &Output & Fairness Post-processing & Re-ranking\cite{karako_using_2018, zehlike_fair:_2017, singh_fairness_2018, patro_fairrec_2020}\\
       &Model/Output&Fairness Certification &Metrics~\cite{kaya:2020}\\
        \hline
\end{tabular}
\caption{Comparison of fairness management methods in the different domains.}
\label{tb:fairness:comparison}
\end{table}

%\section{Explainability Management}\label{sec:explain}
%\input{SECTIONS/explainability_management.tex}

\section{Explainability Management}\label{sec:explain}
%The third step in a comprehensive approach to mitigating algorithmic system bias, is to ensure the transparency of the system by providing a reliable explanation about its process and outcomes.

%\green{The} design of an interpretable system aims to enhance the trust and confidence of the user in the system. Shin and Park~\cite{shin_role_2019} stress the importance of helping the user to understand algorithmic affordances in the adoption and use of a system. They have identified that the user experience is affected by the lack of system transparency and demonstrate statistically that fairness, accountability and transparency in algorithmic systems can help the user to understand how the system takes decisions e.g., recommendations, in turn enhancing their trust in the system.

%Write something of how explainability connects to fairness 
The increasing use of algorithms in decision-making -- especially for critical applications -- has lead to policies requiring clearer accountability for algorithmic decision-making, such as the European Union's General Data Protection Regulation, and its ``Right to Explanation” \cite{goodman_european_2017}. %Explainability in the automated decision-making aims to highlight bias and inequalities in the data and model and also to avoid the unfair use of algorithm’s decision outcome.}
%\green{In general, the goals of explainability in algorithmic} systems are to ensure compatibility with social values such as fairness, privacy, causality, and trust. Completely interpretable models are able to justify the predictions made and search for potential biases or mistakes.} \green{%Most of the fairness tools and techniques described in Section~\ref{sec:fairness} do not address the interpretability issue at the same time. %An exception is the AI Fairness 360 (AIF360)~\cite{bellamy_ai_2018} that provides visualizations of attribute and feature bias localization. However, even if this tool is used for model inspection, it cannot directly used for model or outcome explainability.
Doshi-Velez and Kim~\cite{doshi2017towards} argue that interpretability can help us evaluate if a model is biased or discriminatory by explaining the incompleteness that produces some kind of unquantified bias. On the other hand, Selbst and Barocas~\cite{selbst_intuitive_2018} and Kroll et al.~\cite{kroll_accountable_2016} have demonstrated that even if a model is fully transparent, it might be hard to detect and mitigate bias due to the existence of correlated variables.%To measure the level of interpretability of an algorithmic model, state-of-the-art works use global and local interpretability metrics~\cite{freitas:2014, doshi2017towards, johansson2004truth} such as the model complexity, accuracy and fidelity.

According to Eslami et al.~\cite{eslami_user_2019}, full transparency is neither necessary nor desirable in most systems. One reason is that full transparency may negatively affect users' information privacy~\cite{cheng_mmalfm:_2019}. Moreover, users often need to be provided with details on the decisions made, and not simply with explanations of the outcome. A good example is a qualitative study~\cite{brown_toward_2019} in which participants requested not only information concerning how the algorithm under study took decisions, but also the parameters upon which the decisions were taken.

%requested access to the information, and explanations on, how the algorithm took some of the decisions and the parameters the decisions were taken upon.}
 %They argue that this raises the importance of searching for explainability methods for designing more interpretable systems. 
 
Friedrich and Zanker~\cite{friedrich2011taxonomy} classify explainability into two types: \textit{white-box} and \textit{black-box}. \textit{How} explanations are white-box explanations of the input, output and the process leading to the particular outcome. They provide information focusing on the system's reasoning and data source, which enhances the user satisfaction of the system. \textit{Why} explanations treat the systems as non-transparent and they do not provide any information on how a system works. Instead, they give justifications for outcomes and explain the motivations behind the system, to fill the gap between the user's needs and the system's goals. Rader et al.~\cite{rader_explanations_2018} proposed two additional types of explainability, ``What'' and ``Objective''. \textit{What} explanations only reveal the existence of algorithmic decision-making without providing any additional information on how the system works. This type of explainability aims to raise the users' awareness of the algorithm. \textit{Objective} explains the process of the development of the system and its potential improvement with the objective of preventing or mitigating bias in the system. 

Important aspects for personalized explanations in algorithmic systems include the presentation format of the different types of explanations (e.g., graphical, textual, bullet points), the length of each explanation, and the adopted vocabulary if natural language is used for the explanations. The range of explanations is based on the domain; for example, decisions in the health domain are more critical than in movie recommendations and may need a wider range of explanations of how a system derives its predictions/classifications. Regarding the presentation format, Eiband et al.~\cite{eiband_bringing_2018} proposed a participatory design methodology for incorporating transparency in the design of user interfaces such as to make intelligent systems more transparent and explainable. The process used in the design methodology consists of two main parts. The first part defines the content of an explanation (what to explain) while the second focuses on the presentation format of the explanation (how to explain). 
\iffalse
\begin{figure}[ht]
    \centering
    \includegraphics[width=0.9\textwidth]{explainability_Eiband.png}
    \caption{The stages of the participatory design methodology. The first three stages focus on what to explain in the system (content of an explanation) the last two on how to explain (presentation format). The stages are each guided by central underlying questions and involve different stakeholders~\cite{eiband_bringing_2018}.}
    \label{fig:explain:methodology}
\end{figure}
\fi
In a similar vein, Binnis et al.~\cite{binnis_its_2018} classify a set of explanation styles into four categories based on the type of information they would like to present to the end user: 
\begin{itemize}
    \item \textbf{Input influence style}: A set of input variables are presented to the user along with their positive or negative influence on the outcome.
    \item \textbf{Sensitivity style}: A sensitivity analysis shows how much each of the input values would have to differ in order to change the outcome (e.g., class).
    \item \textbf{Case-based style}: A case from the model's training data that is most similar to the decision outcome is presented to the user.
    \item \textbf{Demographic style}: Using this style, the system presents to the user statistics regarding the outcome classes for people in the same demographic categories as the decision subject, e.g., based on age, gender, income level or occupation.
\end{itemize}

Recent surveys on interpretable machine learning methods and techniques can be found in~\cite{guidotti_survey_2018, angelov_explainable_nodate}. In the following sections, we briefly identify the main explainability approaches used in ML and RecSys systems. 
%Describe general categories of explainability. e highlighting the biases learned by the model
\subsection{Model Explainability}
%Post-hoc explainability techniques are applied to ML models that are not interpretable by design aiming to explain the prediction outcome for each input. The main type of post-hoc explainability methods for explaining the algorithmic model is the \textit{explanations by simplification}.  
%Post-hoc explainability methods can be either \textit{model-agnostic}, applied to any ML model, or \textit{model-specific}. 
Model explainability techniques are primarily used to explain the process of an opaque ML model such as a neural network or a deep learning model. These techniques usually use a transparent model to mimic the model's behavior and be interpretable by humans. %According to~\cite{confalonieri_historical_2021}, they classified into three categories: i) Local Explanations, ii) Global Explanations and iii) Counterfactual Explanations. 
%\subsubsection{Model-agnostic Techniques}
%Model agnostic techniques usually rely on model simplification, local explanations, feature relevance estimation and visualization techniques. 
For instance, some works use a decision tree to mimic the behavior of a non-transparent model such as a neural network~\cite{craven_extracting_1996, krishnan_extracting_1999, boz_extracting_2002, johansson_evolving_2009} and tree ensemble models~\cite{domingos_knowledge_1998, gibbons_cad-mdd:_2013, zhou_learning_2016, schetinin_confident_2007, tan_tree_2016}. The use of decision trees for explaining neural networks was first presented in~\cite{craven_extracting_1996} where the \textit{TREPAN} network implements the algorithmic process of the neural network and returns the representations of the model.  Chipman et al.~\cite{chipman_making_2007} use decision trees as an interpetable predictor model for tree ensemble models by summarizing the forest of trees through clustering, and use the associated clusters as explanation models. A similar technique is the use of decision rules to explain a non-transparent model, for instance, by extracting rules from a trained model such as a neural network (NN), and then using the NN to refine existing rules~\cite{craven_extracting_1996, johansson_evolving_2009, zhou_extracting_2003}.

More recent works use ontologies to represent and integrate knowledge to the model in order to enhance human understandability. An example is the recent extension of TREPAN~\cite{confalonieri_trepan_2020} that uses and integrates knowledge in the form of ontologies in the decision tree extraction to enhance human understandability on decision trees. In addition, Ribeiro et al.~\cite{ribeiro_aligning_2021} use ontologies to explain neural networks (NN). They build small classifiers that map a neural network model’s internal state to concepts from an ontology, enabling the generation of symbolic justifications for the output of NN. An alternative approach has been proposed in~\cite{booth_bayes-trex_nodate}, where the authors present the Bayes-TREX framework, which uses Bayesian inference techniques to explain NN based on the whole dataset, not only the test data. Bayes-TREX takes as input the whole data and finds in-distribution examples that trigger various model behaviors across several contexts. %  can aid model transparency by example across several contexts.

In addition to the aforementioned approaches for explainability of non-transparent ML algorithms, many articles, especially in the domain of recommender systems, propose some approaches for interpreting the ranking (recommender) algorithms. In such systems, the authors aim to provide explanations based on user opinions and evaluation of previous purchases, rather than on the analysis of the ranking algorithm~\cite{wang_explainable_2018}. The aim is to provide personalized explanations by selecting the most appropriate explanation style. Nunes et al.~\cite{nunes_systematic_2017} presented a systematic review on explanations for recommendations in decision support systems where they proposed a taxonomy of concepts that are required for providing explanation. According to Tintarev and Masthoff~\cite{tintarev2007}, there are seven purposes for providing explanations in a recommender system: transparency, scrutability, trust, effectiveness, persuasiveness, satisfaction and efficiency. Park et al.~\cite{park_j-recs_2020} introduce the J-RECS, a recommendation
model-agnostic method that generates personalized justifications based on various types of product and user data (e.g., purchase history and product attributes). Although most of the surveyed works in RecSys provide explanations based on user data, a recent work~\cite{gale_explaining_2020} propose some metrics for measuring explainability and transparency of the ranking algorithm.

A good example that shows the connection of explainability with fairness perception is the recent work of Anik et al.~\cite{anik_data-centric_2021}. In this work, the authors explore the concept of data-centric explanations for ML systems that describe the training data to end users. They first investigate the potential utility of such an approach, including the information about training data that participants find most compelling. In a second study, the authors investigate reactions to the explanations across four different system scenarios. Their results suggest that data-centric explanations have the potential to impact how users judge the trustworthiness of a system and to assist users in assessing fairness. 
\subsection{Outcome (or Post-hoc) Explainability}
Outcome explainability approaches attempt to provide an interpretation for the outcome generated by the model. A recent work focuses on providing both local and pedagogical explanations for the output of ML models \cite{martens2011performance}. Pedagogical explanations are those that teach something about how the model works rather than attempting to represent it directly. Outcome explanations are divided into: \textit{visual explanations}, \textit{local explanations} and \textit{feature relevance explanations} techniques. 

%A general method for explaining the output of a classifier is by using only the input and output of the model to decompose the changes in the algorithm’s prediction outcome into contributions of individual feature values. These contributions correspond to known concepts from coalitional game theory~\cite{strumbelj_efficient_2010}. %Similar approaches applied to complex machine learning models that the logic and output are hard to explain are called ``model-agnostic" approaches that explain the output of any classifier, regardless of the machine learning algorithm used to train it and the type of input data. 
Local explanation approaches are the Local Interpretable Model-Agnostic Explanations (LIME)~\cite{ribeiro_why_2016} and its variations~\cite{guidotti_local_2018, ribeiro_anchors:_2018, turner_model_2016}. The explanations in LIME are only provided through linear models and their respective feature importance. Anchors is another local explanation method proposed by Ribeiro et al.~\cite{ribeiro_anchors:_2018} that uses decision rules for explaining the model sufficiently. %Turner et al.~\cite{turner_model_2016} design the model explanation system (MES) that uses the Monte Carlo algorithm to explain \green{the model's} predictions. 

A post-hoc global explainability method has been proposed in~\cite{balayn_what_2021}. A SEPA framework has been introduced that incorporates post-hoc global explanation methods for image classification tasks. SEPA uses understandable semantic concepts (entities and attributes) that are obtained via crowd-sourcing from local interpretability saliency maps.
 
An example of feature relevant explanation approach is the \textit{ExplainD}, a framework presented in~\cite{poulin_visual_2006} for interpreting the outcome of any non-transparent model. ExplainD uses generative additive models (GAM) to weight the importance of the input features. A unified framework of the class of six existing additive feature importance methods, the SHAP (SHapley Additive exPlanations) has also been introduced in~\cite{lundberg2017unified}. SHAP assigns each feature an importance value for a particular prediction to interpret the predictions.

According to Slack et al.~\cite{slack_fooling_2020}, post-hoc explanation techniques that rely on the input, such as LIME and SHAP, are not reliable since they do not take into consideration the bias in the model. In~\cite{slack_fooling_2020}, the authors proposed a scaffolding technique that scaffolds any biased classifier in a way that its input data remain biased but the generated post-hoc explanations do not reflect the underlying bias.
%A different approach for explaining the outcome of \green{non-transparent} models has been proposed in~\cite{haufe_interpretation_2014}. Haufe et al. transform the non-linear model into a linear interpretable model where the features for a specific prediction are easily to be explained. 

Other examples of feature relevant explanations include the approach used in Horne et al.~\cite{horne_rating_2019} for explaining the spread of fake news and misinformation online. They used an AI assistance framework for providing these explanations to users. This was been shown to improve the user perception of bias and reliability on online news consumption. In another approach, Henelius et al.~\cite{henelius_peek_2014} search for a group of attributes of which the interactions affect the predictive performance of a given classifier, and they evaluate the importance of each group of attributes using the fidelity metric. In addition, Vidovic et al.~\cite{vidovic_feature_2016} propose the measure of feature importance (MFI), which is model-agnostic and can be applied to any type of model. Feature-relevant explanations are also used in~\cite{abuzaid_diff_2018}, where the authors suggest the DIFF operator, a declarative operator that unifies explanation and feature selection queries with relational analytics workloads.

Another widely known category of outcome explainability approaches is the use of counterfactual explanations, which is a special case of feature-related explanations~\cite{sharma_certifai_2020, wang_demonstration_2021}. In~\cite{sharma_certifai_2020}, they propose the CERTIFAI model-agnostic technique that provides counterfactual explanations using a genetic algorithm. The user can use counterfactual explanations to understand the importance of the features. In~\cite{wang_demonstration_2021}, the authors introduce Lewis, an open-source software that provides counterfactual explanations for the decision-making of an algorithm at the global, local and contextual level. For individuals negatively impacted by the algorithm's decision, it provides actionable resources to change the outcome of the algorithm in the future.

Visualization model-specific techniques are used to inspect the training process of a deep neural network (DNN) behavior on images~\cite{bojarski_visualbackprop:_2016, zintgraf_visualizing_2017, xu_show_2015}. In these works, a Saliency Mask (SM) is used as the interpretable local predictor e.g., a part of an image. Similarly, Fong et al.~\cite{fong_interpretable_2017} propose a framework of explanations as meta-predictors for explaining the outcome of deep learning models. The meta-predictor is a rule that predicts the response of the model to certain inputs such as highlighting the salient parts of an image. Another set of works use saliency masks to incorporate the DL network activation into their visualizations~\cite{selvaraju_grad-cam:_2017, zhou_learning_2016, simonyan_deep_2013}.

In RecSys, outcome explainability approaches are used to explain the recommendations to the user. One category of explainability techniques for RecSys are the ones that explain the latent factors that contribute to the decision outcome based on the collection of users' interests and items' characteristics such as Explicit Factor Models~\cite{zhang_incorporating_2015} and Tensor factorization~\cite{chen_this_2018}. Other approaches for explaining recommendations are based on the use of knowledge graphs that relate the items' characteristics and users' behavior, based on their past interactions with the items~\cite{catherine_explainable_2017, HeckelV16}. Visual explanations have recently been used in RecSys to justify the recommendation process in combination with giving more control to the user in a specific context of an interactive social recommender system~\cite{tsai2021effects}. By conducting a user study, the authors investigate how the addition of user control and explainability affect the user perception, user experience, and user engagement. Based on the results, the best user experience happens when there is full explainability and control.

\subsection{Explainability Management Comparison}
Table~\ref{tb:explain} provides a comparison of the solutions focusing on Explainability Management. Explainability approaches have primarily been developed in the context of ML algorithms and systems. The best known methods for explaining the model decision-making process use interpretable models to mimic the behavior of black-box models, i.e., decision trees, decision rules and ontologies. Methods for explaining the decision outcome include feature-relevance, local and global explainability and visualization methods.

There is also a growing literature on explainability within the HCI community. These works suggest that explainability and judgement of the outcome or decision of the system should be provided in order to enhance the trust of the end user in the system. Also in HCI, we found a few works that connect explainability to fairness perception. Finally, explainability approaches have also been widely discussed in RecSys and IR systems. The difference between these approaches and the ones used in ML are that they take into consideration the user's perception and have the specific goal of increasing the trust of the end user in the system. The most popular explainability techniques in the RecSys and IR literature are the visualization methods (outcome explainability) that have been applied to justify the ranking results. 

%\textit{\textbf{[Here we should mention something about explainability and mitigating bias]}}

\begin{table}[ht]
\small
         \centering
         \begin{tabular}{|l|l|l|l|}
         \hline
       \multicolumn{4}{|c|}{\textbf{Explainability Management}}\\\hline
        \textbf{Domain} & \textbf{Problem} & \textbf{Solution Space} & \textbf{Reference(s)} \\\hline
       ML & Model &Model Expainability & Use of decision tree \\ &&&\cite{krishnan_extracting_1999,  domingos_knowledge_1998, gibbons_cad-mdd:_2013, chipman_making_2007, zhou_learning_2016, schetinin_confident_2007, tan_tree_2016}\\
       & Model && Use of decision rules \\&&& \cite{craven_extracting_1996, johansson_evolving_2009,  lu_explanatory_2006}\\
       & Model && Ontologies \cite{confalonieri_trepan_2020, ribeiro_aligning_2021, booth_bayes-trex_nodate}\\
   
      % \hline
       & Output & Outcome Explainability & Local explanations \\&&&\cite{ribeiro_why_2016,  ribeiro_anchors:_2018, turner_model_2016}\\
       &Output/User&& Visualization methods \\&&&\cite{bojarski_visualbackprop:_2016, zintgraf_visualizing_2017, xu_show_2015, fong_interpretable_2017, selvaraju_grad-cam:_2017, zhou_learning_2016, simonyan_deep_2013}\\
       &Output/User&&Counterfactual explanations\cite{sharma_certifai_2020,wang_demonstration_2021}
      \\
      &&& Feature-relevance explanations\cite{henelius_peek_2014, vidovic_feature_2016, slack_fooling_2020, abuzaid_diff_2018}\\
      \hline
       IR&Output/User&Outcome Explainability&Global explanaions\cite{balayn_what_2021}\\
       \hline
       HCI & User/Data& Model Explainability& data-centric explanations\cite{anik_data-centric_2021}\\
       &Output/Data & Outcome Explainability & Feature-relevance explanation \cite{horne_rating_2019}\\
       & User/Output & & Taxonomy of explanations \& Styles\cite{friedrich2011taxonomy, eiband_bringing_2018, binnis_its_2018} \\
       & User/Output && Raise user awareness \cite{rader_explanations_2018}\\
       \hline
       RecSys & Model/User & Model Explainability & Taxonomy of concepts~\cite{nunes_systematic_2017}\\
       &Model/User&& Based on user opinions \cite{wang_explainable_2018, cheng_mmalfm:_2019}\\
       &Output/User&& Personalized explanations \cite{park_j-recs_2020}\\
       &Output/User&&Knowledge graph~\cite{catherine_explainable_2017, heckel_scalable_2017}\\
     
       &Output/User&Output Explainability& Visualization methods \cite{kouki_personalized_2019, verbert2013visualizing, bountouridis_siren:_2019, tsai2021effects}\\
     
       \hline
\end{tabular}
\caption{Comparison of explainability management approaches for the different research domains.}
\label{tb:explain}
\end{table}

\section{Bringing it all together}
\label{sec:disc}
%discussion.tex
Our survey was intentionally broad, as we aimed to provide a ``fish-eye view'' of this complex topic. We did not restrict our review to the literature on fairness and/or discriminatory bias in a social sense; rather, we considered articles describing the problems and solutions surrounding bias, which affect any number of attributes including the quality of information provided by a system.
\iffalse
\paragraph{Stakeholders}
They are important as they play different roles in any of the activities involving bias, e.g., generation, detection, mitigation. From the four classes of stakeholders that we analyzed, Observers, Users and Indirect Users play a different role as compared to Developers, typically working ``outside'' of the system (i.e., having limited or no access to its data and interworkings). They can typically apply only Auditing to the detection of bias, while Developers can also apply Discrimination Detection upon the training data and/or the algorithmic model. Fairness perception and Outcome Explainability mainly concerned Users and Indirect Users. Nonetheless, all stakeholders can play a key role in raising awareness of the problem of algorithmic bias. 
\fi
\begin{figure}[ht]
    \centering
    \includegraphics[width=0.9\textwidth]{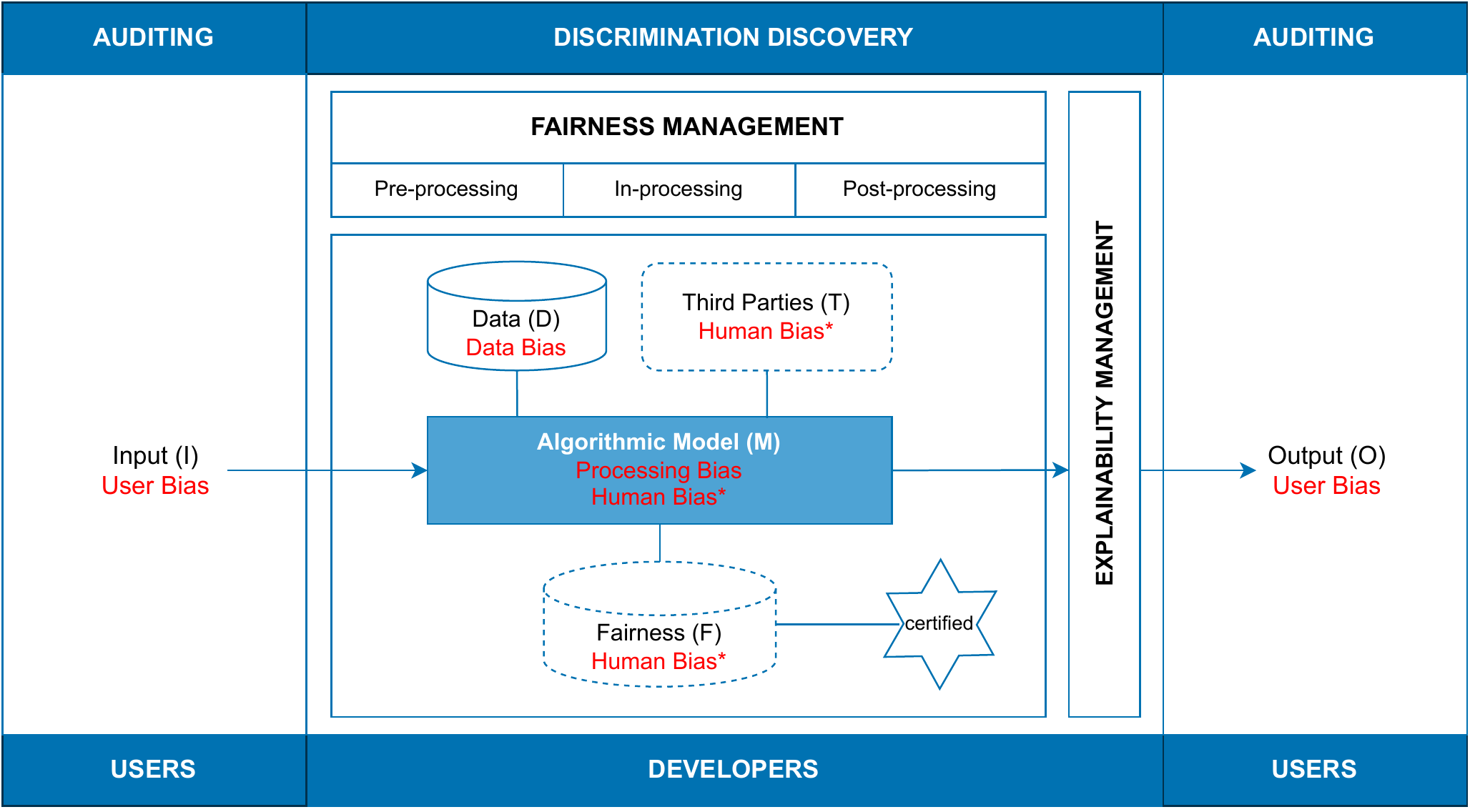}
    \caption{%Observers'
    A fish-eye view of mitigating algorithmic bias: problems, stakeholders, solutions.}
    \label{fig:final}
\end{figure}

\paragraph{Sources of Bias}
Articles reviewed in our survey mentioned at least one of seven problematic components and/or points at which bias can be detected. These are shown in Figure~\ref{fig:final}, which also groups them into four main types: data bias, user bias, processing bias, and human bias. In reality, all biases are at least indirectly \textit{human biases}; for instance, datasets and processing techniques are created by humans. However, we believe that it is helpful to distinguish the biases that are directly introduced into the system by humans, such as third-party biases, those resulting from conflicting fairness constraints, as well as those due to the choices of the developer. User bias is distinguished from other human biases in our framework; as detailed in the literature, users can both introduce bias (e.g., in biased input), but can also perceive bias in the output. Finally, Figure~\ref{fig:final} also incorporates, at a high level, the three steps in a comprehensive solution to mitigate algorithmic bias: bias detection, fairness management and explainability. Figure~\ref{fig:final} presents an overview of the problem and solution spaces revealed by the survey. %, from the point of view of the observer (i.e., researcher). 
This framework integrates the concepts presented earlier on, the components of a system that can be problematic (Figure~\ref{fig:alg_system}), and the solutions described across communities (Figure~\ref{fig:FAT:solutions}).  %including the involvement of multiple stakeholders (Fig.~\ref{fig:roles_bias}),  ).  %Fig.~\ref{fig:venn} presents a Venn diagram showing the potential for cross-fertilization among the four communities that we reviewed, in terms of realizing comprehensive solutions for mitigating bias. The interrelationship between the communities is primarily based on the stakeholders involved in implementing each solution. 

%there are specific solution approaches that are used for all communities such as data re-sampling (fairness pre-processing), discrimination discovery based on specific fairness metrics (or notions). IR and RecSys systems, as ranking systems, use similar solution approaches in all the steps of mitigation bias such as auditing the system by users (bias detection), learning to rank solutions (fairness in-processing), re-ranking solution (fairness post-processing)  and visualization techniques for explaining the decision outcome to the end user (outcome explainability). In HCI community, they use solution approaches for auditing ... }
Next, Table~\ref{tab:cross-fert} depicts the cross-fertilization between the four communities that we reviewed, in terms of realizing comprehensive solutions for mitigating bias. All four communities use all three steps for mitigating bias in different parts of an algorithmic system. However, the interrelationships between the communities is primarily based on the stakeholders involved in implementing each solution. In addition, there are similarities and differences across the specific solution approaches used in each step of mitigating bias within the different communities. For instance, data re-sampling is an approach used in all four communities for fairness pre-processing, while learn-to-rank (fairness in-processing) and re-ranking (fairness post-processing) are approaches used for fairness management only in the communities of IR and RecSys, which are concerned with ranking systems. In the following paragraphs, we give a more detailed view of the interrelationships of the four communities in each step of mitigating bias, considering both the solution approaches and the stakeholders involved in implementing them.

\iffalse
\begin{figure}
    \centering
    \includegraphics[width = 0.55 \textwidth]{SECTIONS/Venn_diagram.PNG}
    \caption{Venn diagram: Cross-fertilization between the four domains.}
    \label{fig:venn}
\end{figure}

\fi

%\paragraph{Cross-fertilization among the communities and stakeholders}
\begin{table}[ht]
  \small
\begin{tabular}{|l|l|l|l|l|}
\hline
        & \multicolumn{4}{c|}{\textbf{Stakeholders}} \\ 
        \hline 
        \textbf{Domains}& \textbf{Developers} & \textbf{Users} & \textbf{Observers} & \textbf{Indirect Users} \\
        \hline
        ML & Bias Detection & & Auditing &\\
        & Fairness management &Fairness Perception&&Fairness Perception\\
        &Model Explainability &Outcome Explainability &&Outcome Explainability\\
        \hline 
        IR & & & Bias Detection&Auditing\\
        &Fairness Pre-processing &Perceived Fairness&Fairness Certification&Perceived Fairness\\
        &Fairness In-processing&&&\\
        &Fairness Post-processing&&&\\
        & &Outcome Explainability && Outcome Explainability\\
        \hline
        RecSys& Bias detection&&Bias Detection&Auditing\\
        &Fairness Pre-processing &Fairness Perception&&Fairness Perception\\
        &Fairness In-processing&&&\\
        &Fairness Post-processing &Fairness Certification&&\\
        &Model Explainability &Outcome Explainability &&Outcome Explainability 
        \\
        \hline
        HCI&&&Bias Detection&Auditing\\
        &&Fairness Perception&Fairness Certicifation&Fairness Perception\\
        &&&Fairness Pre-processing&\\
        &Model Explainability &Outcome Explainability &&Outcome Explainability\\
        \hline
    \end{tabular}
    \caption{Cross-fertilization between research communities.}
    \label{tab:cross-fert}
\end{table}

%\paragraph{Discrimination Detection}
%Discrimination detection is defined as the problem of detecting any form of discrimination bias through the system. The solution approaches, auditing and discrimination discovery have been used in all the four domains that we examined. Both approaches use tools and practices for detecting unfair treatment by data / algorithms / systems.  Discrimination Discovery approaches are used across all domains, and can be applied to the study of particular tasks and/or algorithms (e.g., a top-k ranking algorithm) as well as to deployed systems, which may consist of a whole collection of algorithmic processes (e.g., a proprietary search engine, which uses not only relevance ranking, but also personalization / localization algorithms, among others). The former case is more commonly addressed in the ML literature, while the latter case is more often discussed in domains such as IR and RecSys. 

\textit{Bias Detection:} In most of the articles in our repository, across all four communities, auditing is typically done by the observers of the system. It should also be noted that within ML, beyond involving the model, inputs and outputs, auditing can also involve the generation of biased datasets for conducting a black-box audit. As presented in Table~\ref{tab:cross-fert}, in three domains, bias detection, in general (both auditing and discrimination discovery), is done by observers. The exception is ML, where developers implement automated auditing and discrimination discovery tools, and also observers use auditing to detect fairness issues in the system. In addition, in the RecSys community, developers sometimes implement the auditing process or use discrimination (or fairness) metrics to detect bias as in the ML community.

\textit{Fairness Management:} The issue of ensuring that people and/or groups of people are treated fairly by an algorithmic system was found to be of interest to researchers across all domains considered. However, the tools stakeholders have at their disposal vary. For instance, in three communities (ML, RecSys, IR), developers are the ones who implement pre-processing, in-processing and post-processing methods to mitigate fairness issues in different parts of the algorithmic system, as also presented in Table~\ref{tab:cross-fert}. Specifically, in ML systems, developers are involved both in the development of the system and manage fairness of the system by developing inside the box.
In contrast, in HCI, the system observers manage fairness by observing the system's behavior or the output of the system. In addition, the users of the system participate in the conducted studies for managing system fairness concerning the users' perception. In IR and RecSys, apart from the developers, observers are also involved in certifying that the algorithm is fair.

%Machine learning methods, in general, have been shown to be effective for promoting fairness and transparency in any algorithmic system. Fairness sampling, fairness learning and re-labelling the data are the most commonly used approaches not only in a data mining/ML algorithmic system but also in the HCI, IR and RecSys systems. 

\textit{Explainability Management:}
With respect to the transparency of the algorithmic system, a set of explainability approaches has been introduced in the literature, to encourage trust in the system by the end user, which primarily concerns the HCI and ML communities. In HCI articles, the most appropriate presentation and format of explainability is examined for enriching the transparency of the systems and the trust of the end user. Moreover, multiple papers study specific explainability approaches for explaining the matching/ranking algorithm in RecSys and IR ranking systems. As shown in Table~\ref{tab:cross-fert}, in ML systems, the developers implement algorithms or methods for providing transparency for the black-box model and outcome whereas in RecSys and IR ranking systems, personalized explanations focus on the user and indirect users of the system. In HCI, the observer, in collaboration with the user, conducts experimental studies using various explanation presentation styles and in some cases, personalized explanations for providing the user with some transparency of the system. The exception is one HCI approach, where the developer(s) provides data-centric explanations (Model Explainability).

%some explainability approaches are based on User's opinion where as for some other approaches, the developer of the system has to implement a method for providing transparency of the model and the outcome of the system i.e. matrix factorization.

%While producing models and/or outcomes that are easily interpretable to the user is, in and of itself, viewed as a positive characteristic, it is important to emphasize the particular role of explainability management for bias mitigation. Specifically, in this context, explainability can be viewed as a means rather than an end; complex algorithmic systems can become more transparent to users, the more interpretable their models and outcomes are. Clearly, explainability has a tight relationship to the user's perception of fairness. 

\paragraph{Affected Attributes}

From the articles reviewed in this survey, we can conclude that there are two types of attributes that are affected by the bias and fairness issues in an algorithmic system:

\begin{itemize}
  \item Attributes describing the \textit{social world}; in particular, socio-cultural characteristics of people such as gender, age, language and national origin.
  \item Attributes describing \textit{information}, with the critical question being how well the attributes describe real-world events and phenomena, i.e., the quality and/or credibility of information provided as input to the algorithm, or as output to the user.
%    \item \green{How well attributes describe real world events and phenomena, i.e., the information which is provided to the algorithm before hand.}
\end{itemize}

As mentioned, the attributes describing information are most clearly connected to the explainability management approaches. The other solutions (auditing, discrimination discovery and fairness management) typically address bias that concerns attributes of the real-world and in some cases, information as well. This is the case because in explainability management, people are interested in the process by which information is built while in the other cases, they are interested in the actual discrimination. Based on that, we can also conclude that the three steps of mitigating bias are complementary and can be applied to address different facets of the problem within an algorithmic system.

%Regarding the solutions, independently from the domain that they are applied, in most of the cases, there are more than one stakeholders who are responsible for them. For instance, both developers and observers are responsible for fairness sampling, discrimination discovery and auditing where white-box explainabiliy and black-box outcome explainability.
\paragraph{Limitations}
We must note some challenges faced when reviewing the literature on mitigating algorithmic bias. First, the field is becoming highly interdisciplinary. It was often difficult to categorize the articles we collected into one domain; for instance, RecSys researchers often publish in HCI venues, or even ACM FAccT. Thus, while we aimed to collect articles from across four domains, one should keep in mind that there is some overlap between them. Thus, it was more difficult than expected to characterize how each community has contributed to the work on addressing algorithmic bias. This challenge, however, does not affect the development of a ``fish-eye view'' on the field. In addition, the classifications of solutions that we provide is driven by empirical evidence as we discovered it by the extensive, state-of-the-art works reviewed in the survey. Still, there are cases we do not capture, which are outside of our classification. Any classification scheme has its own foundation issues, which have long-term effects, as they influence the validity of the classification in the long term. Thus, it becomes obvious that the entire issue of bias and the solution(s) to bias should be placed into the context of diversity, taking into account local cultures and problems \cite{Giunchiglia2021}, which will be examined in a future work.

Secondly, the framework presented in Figure~\ref{fig:final} does not yet explicitly incorporate \textit{accountability} into the solutions for mitigating algorithmic bias. Because we focused on literature in the information and computer sciences, studying articles describing particular algorithms and/or systems, the issue of accountability was not often discussed. Going forward, the literature search could be expanded into law and the social sciences as to further investigate the role of the Observer/ Regulator in the landscape of solutions.

%The fish-eye-view review is a first attempt to provide a methodology for mapping the problems, solutions and stakeholders among different communities.}

\section{Conclusion}\label{sec:open}
In this survey, we provided a ``fish-eye view'' of research to date on the mitigation of bias in any type of algorithmic system. With the aim of raising awareness of biases in user-focused, and algorithm-focused systems, we examined studies conducted in four different research communities: information retrieval (IR), human-computer interaction (HCI), recommender systems (RecSys) and machine learning (ML). We outlined a classification of the solutions described in the literature for detecting bias as well as for mitigating the risk of bias and managing fairness in the system. %The main steps for providing discrimination and fairness-aware algorithmic systems are: i) the detection of discrimination bias using Auditing and/or Discrimination discovery, ii) mitigating discrimination and build fairness-aware systems using data mining fairness management approaches, iii) mitigating perceived bias using perceived fairness management approaches, iv) build transparency of the system and enhance the trust of the end user to the system using explainability approaches.
Multiple stakeholders, including the developer (or anyone involved in the pipeline of a system’s development), and various system observers (i.e., stakeholders who are not involved in the development, but who may use, be affected by, oversee, or even regulate the use of the system) are involved in mitigating bias. In future work, we aim to further refine the various roles of individual stakeholders and the relationships between them. 

A second consideration to be explored, is that while many solutions described in the literature have been formalized (e.g., discrimination detection methods, fairness management, internal certification), there are many other issues surrounding \textit{perceived fairness}. The perceived fairness of the user is somewhat subjective and it is not clear how the internal, formal processes relate to users’ perceptions of the systems and their value judgements. To this end, it is important to emphasize the particular role of explainability management for bias mitigation. Specifically, in this context, explainability can be viewed as a means rather than an end; complex algorithmic systems can become more transparent to users, the more interpretable their models and outcomes are. Clearly, explainability has a tight relationship to the user's perception of fairness.

Finally, in this survey, we recorded the attribute(s) affected by the problematic system in each of the reviewed domains and found that there are two key types of attributes affected by the problematic system: attributes describing the world and attributes describing information. Based on that, explainability management solutions mitigate bias that only affects information, while bias detection and fairness management mitigate bias that affects the attributes describing the social world. In future work, we aim to treat the two types of bias (social world, information) independently. %we aim to find connections between these attributes and the solution approaches for detecting and mitigating bias.

%\begin{figure}[h]
%  \centeri
%  \includegraphics[width=\linewidth]{sample-franklin}
%  \caption{1907 Franklin Model D roadster. Photograph by Harris \&
%    Ewing, Inc. [Public domain], via Wikimedia
%    Commons. (\url{https://goo.gl/VLCRBB}).}
%  \Description{The 1907 Franklin Model D roadster.}
%\end{figure}

\begin{acks}
This project is partially funded by the European Union's Horizon 2020 research and innovation programme under grant agreement No. 810105 (CyCAT). Otterbacher and Kleanthous are also supported by the Cyprus Research and Innovation Foundation under grant EXCELLENCE/0918/0086 (DESCANT) and by the European Union's Horizon 2020 Research and Innovation Programme under agreement No. 739578 (RISE).
\end{acks}

\bibliographystyle{ACM-Reference-Format}
\bibliography{bibliography}
\end{document}